\documentclass[preprint,times,12pt]{article}\usepackage{amsmath,amssymb,amsthm}
\usepackage{graphicx}
\usepackage{ifpdf}
\usepackage{color}
\allowdisplaybreaks
\DeclareGraphicsExtensions{.pdf,.png,.jpg,.jpeg,.mps,.eps}

\usepackage{times}
\usepackage[all]{xypic}
\usepackage[colorlinks=false,linktocpage=true]{hyperref}
\usepackage{hyperref}
\hypersetup{colorlinks=false}

\newtheorem{example}{Example}[section]

\newtheorem{remark}{Remark}[section]
\newtheorem{theorem}{Theorem}[section]

\newtheorem{lemma}{Lemma}[section]                                                                                                              

\newtheorem{definition}{Definition}[section]

\numberwithin{equation}{section}

\def\R{\mathbb R}
\def\C{\mathbb C}
\def\N{\mathbb N}
\def\Z{\mathbb Z}
\def\G{\mathsf{G}}
\def\H{\mathbb H}

\def\X{\mathcal X}

\def\P{{\mathcal P}}

\newcommand{\eps}{\varepsilon}
\def\tr{{\rm tr}}
\newcommand{\PSL}{{\rm PSL}}
\newcommand{\SL}{{\rm SL}}

\def\arccosh{{\rm arccosh}}

\def\T{{\mathcal T}}

\begin{document}
	\title{Partner orbits and action differences on compact factors of the hyperbolic plane. Part II: Higher-order encounters}
	\author{{\sc Hien Minh Huynh} \\[1ex] 
		Department of Mathematics,\\
		Quy Nhon University, Binh Dinh, Vietnam;\\
		Universit\"at K\"oln, Institut f\"ur Mathematik \\
		Weyertal 86-90, D\,-\,50931 K\"oln, Germany \\[1ex] 
		e-mail: hhien@mi.uni-koeln.de
		\date{} 
	}
	\maketitle

	\begin{abstract} Physicists have argued that periodic orbit bunching
		leads to universal spectral fluctuations for chaotic quantum systems.
		To establish a more detailed mathematical understanding of this fact, 
		it is first necessary to look more closely at the classical side of the problem 
		and determine orbit pairs consisting of orbits which have similar actions. 
		We specialize to the geodesic flow on compact factors of the hyperbolic plane 
		as a classical chaotic system. 
		The companion paper (Huynh and Kunze, 2015) proved the existence of a unique periodic partner orbit for a given periodic orbit with a small-angle self-crossing in configuration space that is a 2-encounter and  derived an estimate for the action difference of the orbit pair. In this paper,  we provide an inductive argument to deal with higher-order encounters: we prove that a given periodic orbit including an $L$-parallel encounter 
		has $(L-1)!-1$ partner orbits; we construct partner orbits
		and give estimates for the action differences between
		orbit pairs. 
	\end{abstract}
	{\bf Keywords: }
	geodesic flow, periodic solution, partner orbit, higher-order encounter
	
\section{Introduction}
In the semi-classical limit chaotic quantum systems very often exhibit universal behavior,
in the sense that several of their characteristic quantities agree with the respective
quantities found for certain ensembles of random matrices. Via trace formulas, such
quantities can be illustrated as suitable sums over the periodic orbits of the underlying
classical dynamical system. For instance, the two-point correlator function is 
expressed by a double sum over periodic orbits
\begin{equation}\label{formfactor}
K(\tau)=\Big\langle \frac{1}{T_H}\sum_{\gamma,\gamma'}A_\gamma 
A_{\gamma'}^* e^{\frac{i}{\hbar}(S_\gamma-S_{\gamma'})}
\delta\Big(\tau T_H-\frac{T_\gamma+T_{\gamma'}}{2}\Big) \Big\rangle,
\end{equation}
where $\langle\cdot\rangle$ abbreviates the  average over the energy and
over a small time window, $T_H$ denotes the Heisenberg time and $A_\gamma$, $S_\gamma$, and $T_\gamma$ 
are the amplitude, the action, and the period of the orbit $\gamma$, respectively. 

\begin{figure}[ht]
	\begin{center}
		\begin{minipage}{0.8\linewidth}
			\centering
			\includegraphics[angle=0,width=0.8\linewidth]{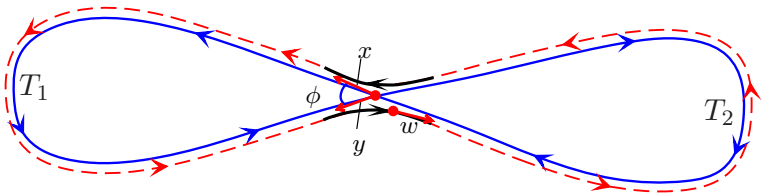}
		\end{minipage}
	\end{center}
	\caption{Example of a Sieber-Richter pair}\label{2encounter}
\end{figure}

The diagonal approximation  $\gamma=\gamma'$ to (\ref{formfactor}) studied by Hannay/Ozorio de Almeida \cite{hoda} 
and Berry \cite{berry} in the 1980's contributes to the first order $2\tau$ to \eqref{formfactor}; see also \cite{KeatRob}.
To next orders, as $\hbar\to 0$, the main term  
from (\ref{formfactor}) arises owing to those orbit pairs $\gamma\neq\gamma'$ 
for which the action difference $S_\gamma-S_{\gamma'}$ is `small'. 
This was first considered by Sieber and Richter \cite{Sieber2,SieberRichter}, 
who predicted that a given periodic orbit with a small-angle self-crossing in configuration space 
will admit a partner orbit with almost the same action. The original orbit and its partner are called  a Sieber-Richter pair. In phase space, a Sieber-Richter pair contains a region where two stretches of each  orbit are almost mutually time-reversed
and one addresses this region as a {\em $2$-encounter} or, more strictly, a {\em $2$-antiparallel encounter};
the `2' stands for two orbit stretches which are close in configuration space, 
and `antiparallel' means that the two stretches have opposite directions (see Figure \ref{2encounter}). It was shown in \cite{SieberRichter} that 
Sieber-Richter pairs contribute to the spectral form factor \eqref{formfactor} the second order term
$-2\tau^2$, and it turned out that the result agreed 
with what is obtained using random matrix theory \cite{efetov}, for certain symmetry classes.  

This discovery prompted an increased research activity on the subject matter 
in the following years and finally led to an expansion 
\[ K(\tau)=2\tau-\tau\,\ln(1+2\tau)=2\tau-2\tau^2+2\tau^3+\ldots \] 
for the orthogonal ensemble (the symmetry class relevant for time-reversal invariant systems) 
to all orders in $\tau$, by including the higher-order encounters also; 
see \cite{HMBH,mueller2009,mueller2004,mueller2005}, and in addition \cite{haake,muellerthesis}, 
which provide much more background and many further references. 

To establish a more detailed mathematical understanding, we start
on the classical side and try to prove the existence of partner orbits and derive good estimates for the action differences 
of the orbit pairs. For $2$-encounters this was done in the previous work \cite{HK}, where we considered 
the geodesic flow on compact factors of the hyperbolic plane; in this case the action of a periodic orbit is proportional to its length. It was shown in \cite{HK} that
a $T$-periodic orbit of the geodesic flow crossing itself in 
configuration space at a time $T_1$ has a unique partner orbit that remains 
$9|\sin(\phi/2)|$-close to the original one and
the action difference between them is approximately equal
$\ln(1-(1+e^{-T_1})(1+e^{-(T-T_1)})\sin^2(\phi/2)))$ with the estimated error 
$12\sin^2(\phi/2)e^{-T}$, where $\phi$ is the crossing angle (see Figure \ref{2encounter}),
and  this proved the accuracy of Sieber/Richter's prediction in \cite{SieberRichter} mentioned above.
In this paper, we continue considering this hyperbolic dynamical system to deal with the technically more involved higher-order encounters. 

In the physics community this system is often called the Hadamard-Gutzwiller model, 
and it has frequently been studied \cite{braun2002,HMBH,Sieber1}; 
further related work includes \cite{haake,mueller2005,Turek05}. 
 For instance, Heusler et al.\;\cite{HMBH} identified the term
 $2\tau^3$ and it was shown that there are 5 families of orbit pairs contributing to $2\tau^3$: 3 families of orbits with two single 2-encounters and 2 families of orbits with 3-encounters. 
 In that way one obtains whole bunches of periodic orbits with controlled and small mutually action difference. Generalizing these
 ideas, in \cite{mueller2004,mueller2005} the expansion of $K(\tau)$ to all orders in $\tau$ could be derived. Here a key point was to consider encounters where more than two orbit stretches are involved;  see also \cite{haake,muellerthesis,Turek05}.  We speak of an {\em $L$-encounter} when $L$ stretches of a periodic orbit are mutually close to each other up to time reversal. 
 For a precise definition one may pick one of $L$ encounter stretches  as a reference 
 and demand that none of the $L-1$ companions be further away than some small distance; see \cite{AltlandBraun}. In other words, 
 all the $L$ stretches must intersect a small Poincar\'e section.
 The orbit enters the {\em encounter region} through {\em entrance ports}
 and leaves it through {\em exit ports}. Using hyperbolicity, 
  one can switch connections between entrance ports and
  exit ports to get new orbits which still remain close to the original one;  and they are called {\em partner orbits}.
  However, not all the connections give genuine periodic orbits since some of them decompose to several shorter orbits (called pseudo-orbits; see \cite{HK}).  M\"uller et al.~\cite{mueller2005} used combinatorics to count the number of partner orbits and provided an approximation for the action difference, but a construction of partner orbits and an error bound of the approximation 
  have not been derived. 
  Furthermore,   it is necessary to arrive a mathematical definition for `encounters', `partner orbits', and to  introduce the notions of `beginning', `end', and `duration' of an encounter. 
\smallskip

The paper is organized as follows. In Section 2, after giving some background material, we  introduce another version of Poincar\'e sections and a respective version of the Anosov closing lemma. In the case that the space is compact, a Poincar\'e section with small radius can be identified with a square, so
every point in a Poincar\'e section can be represented by a unique couple $(u,s)\in\R^2$ 
called {\em unstable} and {\em stable coordinates}. 
In addition, we provide a `connecting lemma' to connect 2 orbits in a pseudo-orbit. In this way, one can construct a genuine periodic orbit  close to a given pseudo-orbit.
In Section 3, after providing mathematically rigorous definitions of 
`encounters', `partner orbits',  `orbit pairs', etc, 
we provide an inductive argument to construct partner orbits for a given orbit 
with a single $L$-parallel encounter. The first step of the inductive argument stated in Theorem \ref{3-encounter}  shows the existence of a unique partner orbit for
a given orbit with a $3$-parallel encounter. The main result of the current paper is Theorem \ref{thmL} which proves that there exist $(L-1)!-1$ partner orbits for a given periodic orbit with an $L$-parallel encounter such that any two piercing points are not too close. 
This condition is expressed by differences of stable and unstable coordinates of the piercing points
which guarantees that the encounter stretches are separated by non-vanishing loops and whence the partner orbits do not coincide. 
 We use combinatorics to count the number of partner orbits and apply the connecting lemma, the Anosov closing lemmas to construct partner orbits. The  
action difference between the orbit pairs can be approximated by terms of coordinates of the piercing points 
with a precisely estimated error.

\medskip

      \noindent 
      {\bf Acknowledgments:} This work was initiated in the framework 
      of the collaborative research program SFB-TR 12 `Symmetries and Universality in Mesoscopic Systems' 
      funded by the DFG, whose financial support is gratefully acknowledged. Special thanks to my supervisor M.~Kunze for his help and support. I enjoyed many fruitful discussions with P.~Braun, K.~Bieder, F.~Haake, G.~Knieper, and S.~M\"uller. 
      
      \setcounter{equation}{0}
\section{Preliminaries}

We consider the geodesic flow on compact Riemann surfaces of constant negative curvature. 
In fact this flow has had a great historical relevance for the development of the whole theory 
of hyperbolic dynamical systems or Anosov systems. It is well-known that any compact orientable surface 
with a metric of constant negative curvature is isometric to a factor $\Gamma\backslash \H^2$, 
where $\H^2=\{z=x+iy\in \C:\, y>0\}$ is the hyperbolic plane endowed  
with the hyperbolic metric $ds^2=\frac{dx^2+dy^2}{y^2}$ and
 $\Gamma $ is a discrete subgroup of the projective Lie group $\PSL(2,\R)=\SL(2,\R)/\{\pm E_2\}$; 
here $\SL(2,\R)$ is the group of all real $2\times 2$ matrices with unity determinant, 
and $E_2$ denotes the unit matrix.  The group $\PSL(2,\R)$  acts transitively on $\H^2$ by 
M\"obius transformations
$z\mapsto \frac{az+b}{cz+d}$.
If the action is free (of fixed points), then the factor $\Gamma\backslash\H^2$  has a Riemann surface structure.
Such a surface is a closed Riemann surface of genus at least $2$ 
and has the hyperbolic plane $\H^2$ as the universal covering. 
The geodesic flow $(\varphi_t^\X)_{t\in \R}$ on the unit tangent bundle $\X=T^1(\Gamma\backslash\H^2)$
goes along the unit speed geodesics on $\Gamma\backslash\H^2$. 
This means that every orbit under 
the geodesic flow $(\varphi_t^\X)_{t\in\R}$ is a geodesic on $\X$ which is the
projection of a geodesic on $\H^2$. 
In addition, every oriented unit speed closed geodesic
  on $\Gamma\backslash\H^2$ is a periodic orbit of the geodesic flow
  $(\varphi_t^\X)_{t\in\R}$ on $\X=T^1(\Gamma\backslash\H^2)$.

  On the other hand, the unit tangent bundle $T^1(\Gamma\backslash\H ^2)$
  is isometric to the quotient space 
  $\Gamma\backslash \PSL(2,\R)=\{\Gamma g,g\in\PSL(2,\R)\}$, 
  which is the system of right co-sets of $\Gamma$ in $\PSL(2,\R)$, by an isometry
  $\Xi$.
  Then  the geodesic flow $(\varphi_t^\X)_{t\in\R}$ can be equivalently described as the natural 
  `quotient flow' $\varphi^X_t(\Gamma g)=\Gamma g a_t$ 
  on $X=\Gamma\backslash\PSL(2,\R)$  associated to the flow $\phi_t(g)=g a_t$ on $\PSL(2,\R)$
  by the conjugate relation 
  \[\varphi_t^\X=\Xi^{-1}\circ\varphi_t^X\circ\Xi\quad \mbox{for all}\quad t\in\R.\]
  Here $a_t\in\PSL(2,\R)$ denotes the equivalence class obtained from the matrix $A_t=\scriptsize\Big(\begin{array}{cc}
  e^{t/2} & 0\\ 0 & e^{-t/2}
  \end{array}\Big)\in\SL(2,\R)$. 
  In fact, there are one-to-one correspondences 
  between the collection of all periodic orbits
  of $(\varphi_t^X)_{t\in\R}$ (denoted by ${\cal PO}_X$),
  the collection of all oriented unit speed closed geodesics
  on $Y$ (denoted by ${\cal CG}_Y$),
  and the conjugacy classes in $\Gamma$ (denoted by ${\mathfrak C}_\Gamma$).
  The period $T$  of a periodic orbit in ${\cal PO}_X$
  and the length $l$ of the corresponding closed geodesic
  in ${\cal CG}_Y$ are related by  $T=l=2\arccosh(\frac{\tr(\gamma)}2)$, where $\gamma$ is
  a representative of the respective  conjugacy class in ${\mathfrak C}_\Gamma$. 
  
  \smallskip 
  There are some more advantages to work on $X=\Gamma\backslash\PSL(2,\R)$
  rather than on $\X=T^1(\Gamma\backslash\H^2)$. One can calculate explicitly the stable and unstable manifolds 
  at a point $x$ to be
  \[W^s_X(x)=\{\theta^X_t(x),t\in\R\}
  \quad \mbox{and}\quad W^u_X(x)=\{\eta^X_t(x), t\in\R\},\]
  where $(\theta^X_t)_{t\in\R}$ and $(\eta^X_t)_{t\in\R}$
  are the {\em horocycle flow} and {\em conjugate horocycle flow}
  defined by $\theta^X_t(\Gamma g)=\Gamma g b_t$ and $\eta^X_t(\Gamma g)=\Gamma g c_t$;  
  here $b_t,c_t\in\PSL(2,\R)$ denote
  the equivalence classes obtained from 
  $B_t=\scriptsize\Big(\begin{array}{cc}1 &t\\ 0&1 \end{array} \big), \ C_t=\scriptsize\Big(\begin{array}{cc}
  1&0\\ t&1
  \end{array}\Big)\in\SL(2,\R)$. 
  The flow $(\varphi^X_t)_{t\in\R}$
  is hyperbolic, that is, 
  for every $x\in X$ there exists an orthogonal and $(\varphi_t^X)_{t\in\R}$-stable splitting of the tangent space
  $T_xX$
  \[T_x X= E^0(x)\oplus E^s(x)\oplus E^u(x)\]
  such that the differential of the flow $(\varphi_t^X)_{t\in \R}$ is uniformly expanding on $E^u(x)$, uniformly contracting on $E^s(x)$ and isometric on $E^0(x)=\langle \frac{d}{dt}\varphi_t^X(x)|_{t=0}\rangle$. Once can choose  
  \begin{eqnarray*}
  	E^s(x)  =  {\rm span}\Big\{\frac{d}{dt}\,\theta^X_t(x)\Big|_{t=0}\Big\}
  	\quad\mbox{and}\quad 
  	E^u(x)  =  {\rm span}\Big\{\frac{d}{dt}\,\eta^X_t(x)\Big|_{t=0}\Big\}.
  \end{eqnarray*}

General references for this section are \cite{bedkeanser,einsward,KatHas}, 
and these works may be consulted for the proofs to all results which are stated above.
In what follows, we will drop the superscript $X$ from $(\varphi^X_t)_{t\in\R}$ to simplify
notation.

\subsection{Poincar\'e sections, stable and unstable coordinates}\label{Poincsec} 
Recall that the Riemann surface $\Gamma\backslash\H^2$ is compact if and only if 
the quotient space  $X=\Gamma\backslash\PSL(2,\R)$ is compact. Then there is $\sigma_0>0$ such that $d_{\G}(g, \gamma g)\ge\sigma_0$ holds 
for all $g\in \G=\PSL(2,\R)$ and $\gamma\in\Gamma\setminus\{e\}$. In the whole paper, we assume that 
$X$ is compact. 
First we recall the definition of Poincar\'e sections in Part I \cite{HK}.
\begin{definition}\label{Poindn1}
Let $x\in X$ and $\eps>0$. The {\em Poincar\'e section} of radius $\eps$ at $x$ is 
\begin{equation*}
  \P_\eps(x)=\{\Gamma(g c_ub_s) : |u|<\eps, |s| <\eps\},
\end{equation*}
where $g\in \G$ is such that $x=\Gamma g$. 
\end{definition}
\begin{lemma} Let $X$ be compact and $\eps\in\, ]0,\sigma_0/4[$. 
For every $y\in \P_\eps(x)$ there exists a unique couple  $(u,s)\in\, ]-\eps,\eps[\,\times\,]-\eps,\eps[$  such that
  \[y=\Gamma(gc_u b_s)\]
  for any $g\in\G$ satisfying $x=\Gamma g$. 
  \end{lemma}
 \noindent {\bf Proof.} By definition such a couple $(u,s)$ does exist.
  To show its uniqueness, suppose that $x=\Gamma g_1=\Gamma g_2$ and $y=\Gamma g_1b_{s_1}c_{u_1}=\Gamma g_2 b_{s_2}c_{u_2}$ for 
  $g_1,g_2\in \G$ and $(u_1,s_1),(u_2,s_2)\in\, ]-\eps,\eps[\,\times\,]-\eps,\eps[$. Then there are $\gamma, \gamma' \in\Gamma$ such that
\[\gamma g_1= g_2\quad\mbox{and}\quad \gamma' g_1c_{u_1}b_{s_1}=g_2c_{u_2}b_{s_2}.\]
Therefore
   \begin{eqnarray*}
   d_{\G}(\gamma^{-1}\gamma' g_1 c_{u_1}, g_1 c_{u_1})
   &=&d_{\G}(\gamma^{-1}g_2c_{u_2}b_{s_2-s_1},g_1c_1)
   =d_{\G}( g_1 c_{u_2}b_{s_2-s_1}, g_1c_{u_1})\\
   &=&d_{\G}( b_{s_2-s_1}, c_{u_1-u_2})
   \leq d_{\G}(b_{s_1-s_2},e)+d_{\G}(c_{u_2-u_1},e)\\
   &\leq& |s_1-s_2|+|u_1-u_2|
   < 2\eps+2\eps<\sigma_0.
   \end{eqnarray*}
From the property of $\sigma_0$, it implies that $\gamma^{-1}\gamma'=e$, so that  $\gamma=\gamma'$. 
Then $g_2c_{u_2}b_{s_2}=\gamma g_1c_{u_1}b_{s_1}=g_2c_{u_1}b_{s_1}$
yields 
 $c_{u_2-u_1}=b_{s_1-s_2}$,
 and consequently $s_1=s_2, u_1=u_2$ by considering matrices.
{\hfill$\Box$}\bigskip

Thus for $\eps\in\,]0,{\sigma_0}/4[$ the mapping
\[\P_\eps(x)\rightarrow\,]-\eps,\eps[\,\times\,]-\eps,\eps[, \ 
y\mapsto (u,s),\]
such that $y=\Gamma(gc_ub_s)$ defines a bijection.

\begin{figure}[ht]
	\begin{center}
		\begin{minipage}{0.7\linewidth}
			\centering
			\includegraphics[angle=0,width=0.6\linewidth]{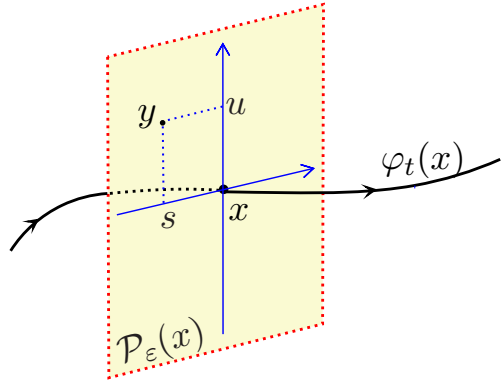}
		\end{minipage}
	\end{center}
	\caption{Coordinatization  of  Poincar\'e section} \label{poi}
\end{figure}

  \begin{definition}\label{deus1}
  The numbers $u=u(y)$ and $s=s(y)$ are called the {\em unstable coordinate} and the {\em stable coordinate} of $y$,
   respectively, and we write $y=(u,s)_x$ (see Figure \ref{poi}).  
\end{definition}
We can also define  Poincar\'e sections as the following.
\begin{definition}\label{pcr2}
Let $x\in X$ and $\eps>0$. The {\em Poincar\'e section} of radius $\eps$ at $x$ is 
\[\P'_\eps(x)=\{\Gamma (gb_sc_u):|s|<\eps,\ |u|<\eps\}\]
where $g\in\G$ is such that $x=\Gamma g$. 
\end{definition}
Note that $b_s$ and $c_u$ are reversed as compared to
$\P_\eps(x)$. We also have the following result. 
\begin{lemma} Let $X$ be compact and $\eps\in\, ]0,\sigma_0/4[$. 
For every $y\in \P'_\eps(x)$ there exists a unique couple  $(s,u)\in\, ]-\eps,\eps[\,\times\,]-\eps,\eps[$  such that
  \[y=\Gamma(gb_sc_u)\]
  for any $g\in\G$ satisfying $x=\Gamma g$. 
  \end{lemma}
 \begin{definition}\label{deus2}
 Again the numbers  $s=s(y)$ and $u=u(y)$ are called the {\em stable coordinate} and the {\em unstable coordinate} of $y$, 
  respectively, and we write $y=(s,u)'_x$. 
\end{definition}

The following result shows that if two trajectories of the flow $(\varphi_t)_{t\in\R}$ intersect a Poincar\'e section and are sufficiently close in the past time (the future time, respectively), then the stable coordinates (unstable coordinates, respectively) of the piercing points are nearly equal. A similar approximation was used in \cite{mueller2005} but left proven.
\begin{theorem}\label{eps0} For every $\rho>0$, there exists  $\eps=\eps(\rho)>0$ satisfying the following property. If $\delta\in \,]0,\frac{\sigma_0}{8}[$ and $x,x_i\in X$ are such that
	$x_i\in \P_\delta(x)$ with $x_i=(u_i,s_i)_{x}$, for $i=1,2$, then for any $T>0$, 
	\begin{itemize}
		\item[(a)]  if $d_X(\varphi_t(x_1),\varphi_t(x_2))<\eps$ for $t\in [0,T]$ then $|u_1-u_2|<\rho e^{-T};$
		\item[(b)] if  $d_X(\varphi_t(x_1),\varphi_t(x_2))<\eps$ for $t\in [-T,0]$ then $|s_1-s_2-s_1s_2(u_1-u_2)|<\rho e^{-T}.$
		\item[(c)] if $d_X(\varphi_t(x_1),\varphi_t(x_2))<\eps$ for $t\in [-T,T]$ then $|u_1-u_2|<\rho e^{-T}$ and $|s_1-s_2|<\frac32\rho e^{-T}.$
	\end{itemize}
\end{theorem}
\noindent{\bf Proof.} Let $\rho>0$ be given.  According to Lemma \ref{konvexa} below, we can select $\eps_1=\eps_1(\rho)>0$ so that $d_{\G}(u,e)<\eps_1$ and $u=\pi(U)\in \G$ for $U=\Big(\scriptsize\begin{array}{cc}u_{11}&u_{12}\\ u_{21}&u_{22}\end{array}\Big)\in\SL(2,\R)
$
yield $|u_{12}|+|u_{21}|<\rho$. Let $\eps=\min\big\{\frac{\sigma_0}{2+\sqrt 2},\eps_1\big\}$.

\smallskip
\noindent
(a)\,\&\,(b) Write $x=\Gamma g$ and 
$x_i=\Gamma g c_{u_i}b_{s_i}$ for $g\in\G$.  Denote $c_i(t)=gc_{u_i}b_{s_i}a_t\in \G$. 
By the definition of $d_X$, for every $t\in [-T,T]$
there is $\gamma(t)\in\Gamma$ so that
$d_X(\varphi_t(x_1),\varphi_t(x_2))=d_{\G}(c_1(t),\gamma(t)c_2(t))<\eps$. Using the property of $\sigma_0$, we can show that $\gamma(t)=\gamma(0)$ for all $t\in [-T,T]$. 
In addition, since $d_{\G}(gc_{u_1}b_{s_1},gc_{u_2}b_{s_2})=d_{\G}(c_{u_1}b_{s_1},c_{u_2}b_{s_2})\leq |u_1|+|u_2|+|s_1|+|s_2|<4\delta<\frac{\sigma_0}{2}$, we have 
\[d_X(\varphi_0(x_1),\varphi_0(x_2))=d_X(x_1,x_2)=
d_{\G}(c_{u_1}b_{s_1},c_{u_2}b_{s_2})=d_{\G}(c_1(0),c_2(0)).\]
This means that we can take $\gamma(0)=e$ and hence
$\gamma(t)=e$  for all $t\in [-T,T]$.  Then $d_{\G}(c_1(t),c_2(t))<\eps$ for all $t\in [-T,T]$, i.e.,
\[d_{\G}(c_{u_1}b_{s_1}a_t,c_{u_2}b_{s_2}a_t)<\eps\leq\eps_1\quad \mbox{for all}\quad t\in[-T,T],\]
or equivalently,
\[d_{\G}(a_{-t}b_{-s_2}c_{u_1-u_2}b_{s_1}a_t,e)<\eps_1\quad \mbox{for all}\quad t\in[-T,T].\]
Write $a_{-t}b_{-s_2}c_{u_1-u_2}b_{s_1}a_t=[H(t)]$
with
\[H(t)=\bigg(\begin{array}{cc}1-s_2(u_1-u_2)&e^{-t}(s_1-s_2-s_1s_2(u_1-u_2))\\
e^t(u_1-u_2)&1+s_1(u_1-u_2)
\end{array}\bigg)\in\SL(2,\R).\]
Whence the definition of $\eps_1$ leads to
\[e^t|u_1-u_2|+e^{-t}|s_1-s_2-s_1s_2(u_1-u_2)|<\rho \quad\mbox{for}\quad t\in [-T,T].\]
In particular, $t=T$ and $t=-T$ imply (a) and (b), respectively.
\smallskip

\noindent
(c) This follows from (a) and (b).
{\hfill$\Box$}\bigskip

We also have a reverse statement.
\begin{theorem} For $\eps\in\,]0,\frac{\sigma_0}4[$. Assume that $x,x_1,x_2\in X$ are such that $x_i\in \P_{\frac{\eps}5}(x)$
	and $x_i=(u_i,s_i)_x$ for $i=1,2$. Then
	for any $T>0$,
	\begin{itemize}
		\item[(a)] if $|u_1-u_2|<\frac\eps 2 e^{-T}$ then $d_X(\varphi_t(x_1),\varphi_t(x_2))<\eps$\ for all\ $t\in [0,T]$;
		\item[(b)] if $|s_1-s_2-s_1s_2(u_1-u_2)|<\frac\eps2 e^{-T}$ then $d_X(\varphi_t(x_1),\varphi_t(x_2))<\eps$\ for all \ $t\in [-T,0]$;
		\item[(c)] if $|u_1-u_2|+|s_1-s_2|<\frac{\eps}{2}e^{-T} $ then $ d_X(\varphi_t(x_1),\varphi_t(x_2))<\eps$\ for all\ $t\in [-T,T].$
	\end{itemize}
\end{theorem}
\noindent{\bf Proof.}
(a) Let $T>0$ be given  and fix $t\in [0,T]$. We have
\begin{eqnarray*}
	d_X(\varphi_t(x_1),\varphi_t(x_2))
	&=& d_X(\Gamma gc_{u_1}b_{s_1}a_t,\Gamma gc_{u_2}b_{s_2}a_t)
	\leq d_\G(gc_{u_1}b_{s_1}a_t, gc_{u_2}b_{s_2}a_t)\\
	&=& d_\G(c_{(u_1-u_2)e^{t}}b_{s_1e^{-t}},b_{s_2e^{-t}})
	\leq d_\G(c_{(u_1-u_2)e^{t}}b_{s_1e^{-t}},e)+d_\G(b_{s_2e^{-t}},e)\\
	&\leq& d_\G(c_{(u_1-u_2)e^{t}},e)+d_\G(b_{s_1e^{-t}},e)+d_\G(b_{s_2e^{-t}},e)
	\leq |u_1-u_2|e^{t}+(|s_1|+|s_2|)e^{-t}\\
	&<&\frac\eps2 e^{t-T}+2\frac\eps5 e^{-t}
	<\eps.
\end{eqnarray*}
(b) First, we write $c_{u_i}b_{s_i}=b_{\tilde s_i}c_{\tilde u_i}a_{\tilde\tau_i}$ with
\[ \tilde\tau_i=-2\ln(1+u_is_i),\quad \tilde u_i=u_i(1+u_is_i),\quad \tilde s_i=\frac{s_i}{1+u_is_i}. \]
For $t\in [-T,0]$, analogously to (a), we obtain
\begin{eqnarray*}
	d_X(\varphi_t(x_1),\varphi_t(x_2))
	&\leq& d_\G(c_{u_1}b_{s_1}a_t, c_{u_2}b_{s_2}a_t)
	= d_\G(b_{\tilde s_1}c_{\tilde u_1}a_{\tilde\tau_1}a_t,b_{\tilde s_2}c_{\tilde u_2}a_{\tilde\tau_2}a_t)\\
	&=& d_\G(b_{(\tilde s_1-\tilde s_2)e^{-t}}c_{\tilde u_1e^t}a_{\tilde\tau_1},c_{\tilde u_2e^t}a_{\tilde\tau_2})\\
	&\leq&d_\G(b_{(\tilde s_1-\tilde s_2)e^{-t}},e)+d_\G(c_{\tilde u_1e^t},e)+d_\G(c_{\tilde u_2e^t},e)+d_\G(a_{\tilde\tau_1},e)+d_\G(a_{\tilde\tau_2},e)\\
	&\leq&|\tilde s_1-\tilde s_2|e^{-t}+(|\tilde u_1|+|\tilde u_2|)e^t+\frac1{\sqrt2}(|\tilde\tau_1|+|\tilde\tau_2|)\\
	&\leq&|s_1-s_2-s_1s_2(u_1-u_2)|e^{-t}+(|\tilde u_1|+|\tilde u_2|)e^t+\frac4{\sqrt2}(|u_1s_1|+|u_2s_2|) \\
	&<&\frac\eps2 e^{-T-t}+\frac{2\eps}5+\frac{4}{\sqrt 2}\cdot\frac{2\eps^2}{25}
	<\eps.
\end{eqnarray*}
(c) It follows from $|u_1-u_2|+|s_1-s_2|<\frac{\eps}{2}e^{-T}$ that
$|u_1-u_2|<\frac{\eps}{2}e^{-T}$ and
$|s_1-s_2-s_1s_2(u_1-u_2)|<\frac{\eps}{2}e^{-T}$ which prove (c) by (a) and (b).
{\hfill$\Box$}\bigskip

\subsection{Shadowing lemma, Anosov closing lemmas, and connecting lemma}
We recall the shadowing lemma and reformulate the Anosov closing lemma in Part I \cite{HK};

Denote by $W^s_{X,\,\eps}(x)=\{\Gamma(gb_t): |t|<\eps\}$
and
$W^u_{X,\,\eps}(x)=\{\Gamma(gc_t): |t|<\eps\}$ 		 
for $x=\Gamma g$  the {local stable} and {local unstable manifold} 
of $x$ of size $\eps$, respectively. 
\begin{theorem}[Shadowing lemma]\label{shadlemII} 
		If $\eps>0$, $x_1, x_2\in X$, and $x\in W^s_{X,\,\eps}(x_1)\cap W^u_{X,\,\eps}(x_2)$, then 
	\[ d_X(\varphi_t(x_1), \varphi_t(x))<\eps e^{-t}
	\quad\mbox{for all}\quad t\in [0, \infty[ \]
	and 
	\[ d_X(\varphi_t(x_2), \varphi_t(x))<\eps e^t 
	\quad\mbox{for all}\quad t\in\, ]-\infty, 0]. \]
\end{theorem}
\begin{figure}[ht]
	\begin{center}
		\begin{minipage}{\linewidth}
			\centering
			\includegraphics[angle=0,width=0.6\linewidth]{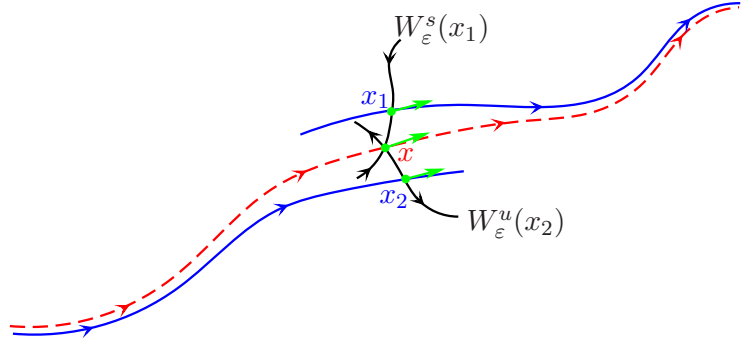}
		\end{minipage}
	\end{center}
	\caption{ Shadowing lemma}\label{Anosovlemma}
\end{figure}

\begin{theorem}[{Anosov closing lemma I}]\label{anosov1}
Suppose that $\eps\in\, ]0, \min\{\frac{1}{4}, \frac{\sigma_0}{8}\}[$, 
$x\in X$, $T\ge 1$, and $\varphi_T(x)\in {\cal P}_\eps(x)$. Let $\varphi_T(x)={(u, s)}_x$, 
in the notation from Definition \ref{deus1}. Then there are $x'\in {\cal P}_{2\eps}(x)$ with $x'=(\sigma,\eta)_x$ and $T'\in\R$ so that 
\begin{equation*}\label{x'}\varphi_{T'}(x')=x'\quad \mbox{and}\quad d_X(\varphi_t(x),\varphi_t(x'))<2|u|+|\eta|< 4\eps\quad \mbox{for all}\quad t\in[0,T].
\end{equation*}
Furthermore,
\begin{equation*}\label{T-T'1}
\Big|\frac{T'-T}2-\ln(1+us)\Big| <5|us|e^{-T},
\end{equation*}
\begin{equation*}\label{TT'anosov1}
e^{T'/2}+e^{- T'/2}=e^{T/2}+e^{-T/2}+us e^{T/2},
\end{equation*}
and
\begin{eqnarray*}
|\sigma|<2|u|e^{-T},\quad |\eta-s|<2s^2|u|+2|s|e^{-T}.
\end{eqnarray*}
\end{theorem}

Using the version of Poincar\'e sections in Definition \ref{pcr2}, we have a respective  statement for the Anosov closing lemma which will be also useful afterwards.

\begin{theorem}[{Anosov closing lemma II}\empty]\label{anosov2}
Suppose that $\eps\in\, ]0, \min\{\frac{1}{4}, \frac{\sigma_0}{8}\}[$, 
$x\in X$, $T\ge 1$, and $\varphi_T(x)\in \P'_\eps(x)$. Let $\varphi_T(x)={(s, u)}'_x$, 
in the notation from Definition \ref{deus2}. Then there are $x'\in {\cal P}'_{2\eps}(x)$ with $x'=(\eta,\sigma)'_x$ and $T'\in\R$ so that 
\begin{eqnarray}\label{periodicx2}\varphi_{T'}( x)= x\quad \mbox{and}\quad d_X(\varphi_t(x),\varphi_t(x'))\leq 2|u|+|\eta|<4\eps\quad \mbox{for all}\quad t\in[0,T].
\end{eqnarray}
Furthermore,
	\begin{equation}\label{T'-Tanosov2}
\Big|\frac{T'}2-\frac{T}2\Big| <4|us|e^{-T},
\end{equation}
\begin{equation}\label{TT'anosov2}
e^{T'/2}+e^{-T'/2}=e^{T/2}+e^{-T/2}+us e^{-T/2},
\end{equation}
and
\begin{equation}\label{etasigmaII}
|\eta-s|\leq 2|s|e^{-T},\quad |\sigma|<2|u|e^{-T}.
\end{equation}
\end{theorem}
\begin{figure}[h]
	\begin{center}
		\begin{minipage}{\linewidth}
			\centering
			\includegraphics[width=0.5\linewidth]{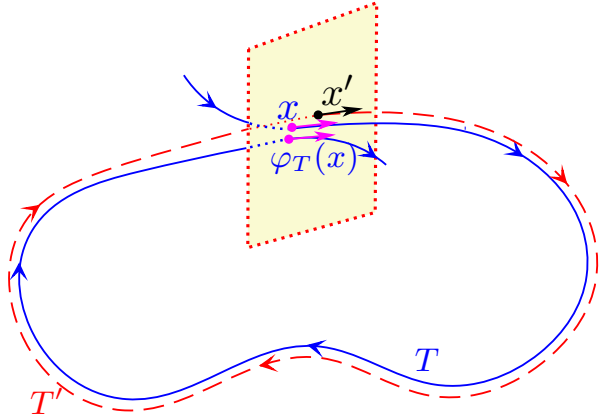}
		\end{minipage}
	\end{center}
	\caption{ Anosov closing lemma}\label{Anosovlemma}
\end{figure}
\noindent
{\bf Proof.} See Figure \ref{Anosovlemma} for an illustration.
By assumption, there are $s,u\in \,]-\eps,\eps[$ and $g\in\G$ such that $\Gamma g=x$ and
$ \varphi_T(x)= \Gamma gb_sc_u.$
Then there is $\zeta\in\Gamma$ such that
$\zeta ga_T= gb_sc_u$ or $\zeta=gb_sc_ua_{-T}g^{-1}.$
The equation 
\begin{equation*}\label{etaeqn}
 ue^{-T}\eta^2-((1+su)e^{-T}-1)\eta-s=0
\end{equation*}
has a solution $\eta\in\R$ satisfying 
$|\eta-s|<2|s|e^{-T}$ as well as $|\eta|<2|s|$.
Then 
\begin{equation*}
\sigma:=\frac{u}{1+(s-\eta)u-\eta u-e^T}
\end{equation*}
is well-defined  and 
$|\sigma|<2|u| e^{-T}.$ Put $g'=gb_\eta c_\sigma\in \G$ and $x'=\Gamma g'$
to obtain $x'\in\P'_{2\eps}(x)$. Defining
\begin{equation}\label{T',anoso2} T'=T-2\ln(1+(s-\eta)u),
\end{equation}
we have 
\begin{equation*} 
\Big|\frac{T'}2-\frac{T}2\Big|=|\ln(1+(s-\eta)u)|\leq 2|s-\eta||u|< 4|us|e^{-T},
\end{equation*}
which is \eqref{T'-Tanosov2}.
Similarly to the proof of the Anosov closing lemma I, we can check that
$\zeta g b_\eta c_\sigma a_{T'}=gb_\eta c_\sigma$
and hence $\varphi_{T'}(x')=x'$ and we also have the latter of \eqref{periodicx2}. Due to
 $\zeta=gb_sc_ua_{-T}g^{-1}=gb_\eta c_\sigma a_{-T'}c_{-\sigma}b_{-\eta}g^{-1}$, this implies  \eqref{TT'anosov2}; and \eqref{etasigmaII} can be done analogously to the Anosov closing lemma I. 
{\hfill$\Box$}\bigskip

\begin{lemma}[{Connecting lemma}\hskip0cm]\label{connect0}
 Let $x_j\in X$ be a $T_j$-periodic point of the flow $(\varphi_t)_{t\in\R}$, for $j=1,2$ and $T_1+T_2\geq 1$ and let $\eps>0$.
If $x_2\in \P_{\eps}(x_1)$, then there is a periodic orbit
$5\eps$-close to the orbits of $x_1$ and $x_2$. More precisely, 
if $x_1=\Gamma g_1$ and $x_2=\Gamma g_1c_ub_s$, then there are
$x\in X$ and $T>0$ such that $\varphi_T(x)=x$,
\begin{eqnarray*}
d_X(\varphi_t(x),\varphi_t(x_1))& < &5\eps\quad \mbox{for all}\quad t\in[0,T_1],\\
d_X(\varphi_{t+T_1}(x),\varphi_t(x_2))&<&5\eps\quad \mbox{for all}\quad t\in[0,T_2],
\end{eqnarray*}
and
\begin{equation}
 \Big|\frac{T-(T_1+T_2)}{2}-\ln(1+us)\Big| <3|us|(e^{-T_1}+e^{-T_2}) +8|us|e^{-T_1-T_2}.
\end{equation}
Furthermore, $x=\Gamma g_1 c_{u e^{-T_1}+\sigma} b_\eta$, where $\sigma,\eta\in\R$ satisfy
\begin{equation}
|\eta-s|<2s^2|u|+2|s|e^{-T_1-T_2}\quad \mbox{and}\quad
|\sigma|<2|u| e^{-T_1-T_2}.
\end{equation} 
\end{lemma}
\begin{figure}[ht]
	\begin{center}
		\begin{minipage}{\linewidth}
			\centering
			\includegraphics[angle=0,width=0.5\linewidth]{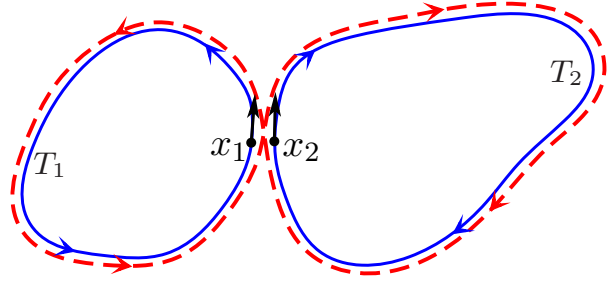}
		\end{minipage}
	\end{center}
	\caption{Reconnection a pseudo-orbit yields a genuine periodic orbit}\label{connect}
\end{figure}

\noindent{\bf Proof.} See Figure \ref{connect} for an illustration. 
Write $x_j=\Gamma g_j$ with $g_j\in\G$. Then
$\Gamma g_2=\Gamma g_1 c_ub_s$ and hence 
$\Gamma g_2b_{-s}=\Gamma g_1c_u=:w \in W^s_{\eps}(x_2)\cap W^u_{\eps}(x_1)$.
By the shadowing lemma (Theorem \ref{shadlemII}),
\begin{equation}\label{t>0}
 d_X(\varphi_t(w),\varphi_t( x_2))<\eps e^{-t},\quad\mbox{for all}\quad t\geq 0
\end{equation}
and
\begin{equation}\label{t<0}
 d_X(\varphi_t(w),\varphi_t( x_1))<\eps e^t,\quad\mbox{for all}\quad t\leq 0.
\end{equation}
For $\hat w=\varphi_{-T_1}(w)$, we verify that
$\varphi_{T_1+T_2}(\hat w)\in\P_{2\eps}(\hat w)$. 
Indeed, 
\begin{eqnarray*}
\varphi_{T_1+T_2}(\hat w)
&=&\Gamma g_2 b_{-s}a_{T_2}
=\Gamma g_2b_{-se^{-T_2}}
=\Gamma g_1c_ub_{s(1-e^{-T_2})}
=\Gamma g_1c_uc_{-u}a_{-T_1}c_ub_{s(1-e^{-T_2})}\\
&=&\Gamma g_1c_ua_{-T_1}c_{u(1-e^{-T_1})}b_{s(1-e^{-T_2})}
=\Gamma (g_1c_u a_{-T_1}) c_{\tilde u}b_{\tilde s}\in \P_{\eps}(\hat w), 
\end{eqnarray*}
where 
\begin{equation}\label{tildeutildes}
\tilde u=u(1-e^{-T_1})\quad \mbox{and}\quad \tilde s=s(1-e^{-T_2}).
\end{equation}
By the assumption $T_1+T_2\ge 1$, we apply the Anosov closing lemma I to get \\
$x=\Gamma (g_1 c_ua_{-T_1})c_\sigma b_\eta \in \P_{2\eps}(\hat w)$ and $T\in\R$ such that
$\varphi_T(x)=x$,
\begin{eqnarray}\label{x}
d_X(\varphi_t(x),\varphi_t(\hat w))<4\eps\quad \mbox{for all}\quad t\in [0,T_1+T_2]
\end{eqnarray}
and
\begin{equation}
\Big|\frac{T-(T_1+T_2)}2-\ln(1+\tilde u\tilde s)\Big|\leq 5|\tilde u\tilde s| e^{-T_1-T_2}.
\end{equation}
For $t\in [0,T_1],$ by (\ref{x}) and (\ref{t<0})
\begin{eqnarray*}
 d_X(\varphi_t(x),\varphi_t(x_1))
& \leq & d_X(\varphi_t(x),\varphi_t(\hat w))+d_X(\varphi_t(\hat w),\varphi_t(x_1))\\
&=& d_X(\varphi_t(x),\varphi_t(\hat w))+d_X(\varphi_{t-T_1}(w),\varphi_{t-T_1}(x_1))<5\eps.
\end{eqnarray*}
For $t\in[0,T_2]$, by (\ref{x}) and (\ref{t>0})
\begin{eqnarray*}
 d_X(\varphi_{t+T_1}(x),\varphi_t(x_2))
&\leq&d_X(\varphi_{t+T_1}(x),\varphi_{t+T_1}(\hat w))+d_X(\varphi_{t+T_1}(\hat w),\varphi_t(x_2))\\
&=& d_X(\varphi_{t+T_1}(x),\varphi_{t+T_1}(\hat w))+d_X(\varphi_{t}(w),\varphi_{t}(x_2))<5\eps.
\end{eqnarray*}
Furthermore,
\begin{eqnarray*}
\Big|\frac{T-(T_1+T_2)}{2}-\ln(1+us)\Big|
& \leq &
 \Big|\frac{T-(T_1+T_2)}{2}-\ln(1+\tilde u\tilde s)\Big| +|\ln(1+\tilde u\tilde s)-\ln(1+us)|
\\
&\leq& 5|\tilde u\tilde s|e^{-T_1-T_2}
+3|\tilde u\tilde s-us| =  3|us|(e^{-T_1}+e^{-T_2})+8|us|e^{-T_1-T_2},
\end{eqnarray*}
due to \eqref{tildeutildes}. 
Finally, since $c_ua_{-T_1}=a_{-T_1}c_{ue^{-T_1}}$ and $x_1$ is $T_1$-periodic, we obtain
\begin{equation*} 
x=\Gamma g_1c_ua_{-T_1}c_\sigma b_\eta=\Gamma g_1 a_{-T_1}c_{u e^{-T_1}}c_\sigma b_\eta=\Gamma g_1c_{ue^{-T_1}+\sigma}b_\eta,
\end{equation*}
where
\[
|\sigma|<2|\tilde u|e^{-T_1-T_2}<2|u|e^{-T_1-T_2}\]
and
\[
|\eta-s|<|\eta-\tilde s|+|\tilde s-s|< 2s^2|u|+2|s|e^{-T_1-T_2}\]
which completes the proof.
{\hfill$\Box$}\bigskip

\subsection{Some auxiliary results}

\begin{lemma}\label{konvexa}
	For every $\eps>0$ there is $\delta>0$ with the following property. 
	If $d_{\PSL(2,\R)}(g,h)<\delta$ then there are
	\[ G=\bigg(\begin{array}{cc}g_{11}&g_{12}\\
	g_{21}&g_{22}
	\end{array}\bigg)\quad \mbox{and}\quad 
	H=\bigg(\begin{array}{cc}h_{11}&h_{12}\\
	h_{21}&h_{22}\end{array}\bigg) \]
	such that $g=\pi(G), h=\pi(H)$ and
	$|g_{11}-h_{11}|+|g_{12}-h_{12}|+|g_{21}-h_{21}|+|g_{22}-h_{22}|<\eps$.
\end{lemma}
See Lemma 2.17\,(b) in \cite{HK} for a proof.
Using the decomposition $g=c_ub_sa_\tau$, we have the following result. 
\begin{lemma}\label{bcbcbc}
If $s_1,s_2,s_3,u_1,u_2\in\, ]-\frac16,\frac16[$, then 
$b_{s_1}c_{u_1}b_{s_2}c_{u_2}b_{s_3}=c_ub_sa_\tau$
 for
 \begin{eqnarray*}
 u &=&
 u_1+u_2+\frac{u_1u_2s_2-(u_1+u_2)\rho}{1+\rho},
 \\
 s &=&
s_1+s_2+s_3+\rho ((2+\rho)s_3+s_1+s_2)
 +u_1s_1s_2(1+\rho),
 \\
 \tau &=& 2\ln(1+\rho),
 \end{eqnarray*}
 where \[\rho=u_2(s_1+s_2)+u_1s_1(1+u_2s_2).\]
\end{lemma}


  \medskip 
  
\begin{lemma}\label{lm1} Let $\eps\in\,]0,\min\{\frac12,\frac{\sigma_0}{12}\}[$ and  $x,x_1,x_2\in X$ be given. 
	If $x_j\in\P_\eps(x)$ and $x_j=(u_j,s_j)_x, j=1,2$, then there is $\tau\in\R$
	such that $\varphi_\tau(x_2)\in \P_{3\eps}(x_1)$. Furthermore,
	$\varphi_\tau(x_2)=(u,s)_{x_1}$ satisfies
	\begin{eqnarray*}
		|u-(u_2-u_1)|<8\eps^3,\quad |s-(s_2-s_1)|<8\eps^3,\quad |\tau|<8\eps^2,
	\end{eqnarray*}
	and \[
	|us-(u_2-u_1)(s_2-s_1)|=|s_1s_2|(u_2-u_1)^2.\]
\end{lemma}
\noindent{\bf Proof.}
Write $x=\Gamma g, x_1=\Gamma g_1, x_2=\Gamma g_2$
for $g,g_1,g_2\in\G$. 
By assumption, $\Gamma g_j=\Gamma g c_{u_j}b_{s_j}$ for $j=1,2$. Then  $\Gamma g_1b_{-s_1}c_{u_2-u_1}b_{s_2}=\Gamma g_2$, and $u,s,\tau$ are the numbers satisfying the decomposition
$b_{-s_1}c_{u_2-u_1}b_{s_2}=c_ub_sa_{-\tau}$.
{\hfill$\Box$}\bigskip

\begin{lemma}\label{lemT-T'}
Let $\eps,T, T'>0$. If
\begin{equation*}
d_X(\varphi_t(x_1),\varphi_t(x_2))<\eps\quad\mbox{for all}\quad t\in [0,\min\{T,T'\}],
\end{equation*}
then 
\begin{equation*}
d_X(\varphi_t(x_1),\varphi_t(x_2))<\eps+\sqrt{2}\,|T-T'|\quad\mbox{for all}\quad t\in [0,\max\{T,T'\}].
\end{equation*} 
\end{lemma}
\noindent{\bf Proof.}
Write $x_1=\Gamma g_1 $ and $x_2=\Gamma g_2$. Without loss of generality, we can assume that
 $T\leq T'$. By assumption,  for $t\in [0, T]$,
\begin{equation*}
d_X(\varphi_t(x_1),\varphi_t(x_2))<\eps\leq \eps+\sqrt{2}\,|T-T'|.
\end{equation*} 
For $t\in [T,\max\{T,T'\}]=[T,T']$, 
\begin{eqnarray*}
d_X(\varphi_t(x_1),\varphi_t(x_2))
&\leq& d_X(\varphi_t(x_1),\varphi_T(x_1))
+d_X(\varphi_T(x_1),\varphi_T(x_2))+d_X(\varphi_T(x_2),\varphi_t(x_2))
\\
&=&d_X(\Gamma g_1a_t,\Gamma g_1a_T)+ d_X(\varphi_T(x_1),\varphi_T(x_2))
+d_X(\Gamma g_2a_t,\Gamma g_2a_T)
\\
&\leq&\eps+ 2d_{\G}(a_t,a_T)=\eps+\sqrt{2}\,|t-T|
\leq \eps+\sqrt{2}\,|T'-T|.
\end{eqnarray*}
{\hfill$\Box$}

Any periodic orbit of the flow $(\varphi_t)_{t\in\R}$ never comes back to another point on the stable manifold or the unstable manifold of a point on it. This follows from the next result.

\begin{lemma}\label{nocom} Assume that $x,y\in X$ are periodic points of the flow $(\varphi_t)_{t\in\R}$ with the same period. Then
	$y\notin W_X^s(x)$ and $y\notin W_X^u(x)$.
\end{lemma}

\noindent 
{\bf Proof.}
Let $x,y$ be $T$-periodic points and suppose on the contrary that
$y\in W_X^s(x), y\ne x$. 
But then 
\[d_X(x,y)=d_X(\varphi_{mT}(x),\varphi_{mT}(y))
\rightarrow 0\quad \mbox{as}\quad m\rightarrow \infty\]
which is impossible since $d_X(x,y)>0$.  
The case of unstable manifold can be treated analogously.
{\hfill$\Box$}\bigskip

Owing to the hyperbolicity two periodic orbits with similar periods 
cannot stay too close together without being identical; see \cite{HK} for a proof.

\begin{lemma}\label{per-coinc} Let $X=\Gamma\backslash {\rm PSL}(2, \R)$ be compact. 
	Then there is $\eps_\ast>0$ with the following property. 
	If $\eps\in \,]0, \eps_\ast[$ and if $x_1, x_2\in X$ 
	are periodic points of ${(\varphi_t)}_{t\in\R}$ having the periods $T_1, T_2>0$ 
	such that $|T_1-T_2|\le\sqrt 2\eps$ and 
	\[ d_X(\varphi_t(x_1), \varphi_t(x_2))<\eps
	\quad\mbox{for all}\quad t\in [0, \min\{T_1, T_2\}], \]  
	then the orbits of $x_1$ and $x_2$ under ${(\varphi_t)}_{t\in\R}$ are identical. 
\end{lemma} 

\section{Higher-order encounters}

In this section, we provide rigorous definitions of `$L$-encounters' and `partner orbits',
 then we give an inductive argument to prove that a periodic orbit involving an $L$-parallel encounter satisfying a certain condition has $(L-1)!-1$ genuine 
 partner orbits, and we derive an estimate for the action difference. 

\subsection{Encounters and partner orbits}\label{ensec}
We continue denoting  $\X=T^1(\Gamma\backslash\H^2)$ and  $X=\Gamma\backslash\PSL(2,\R)$.
\begin{definition}[Time reversal]
The {\em time reversal map} $\T:\X \rightarrow \X$ is defined by 
\[\T(p,\xi)=(-p,\xi)\quad \mbox{for}\quad (p,\xi)\in \X.\]
The respective time reversal map on $X$ is determined by 
 \begin{equation*}\label{revedn}
 \T(x)=\Gamma gd_\pi \quad \mbox{for}\quad  x=\Gamma g\in X,
 \end{equation*}
 where $d_\pi\in\PSL(2,\R)$ is the equivalence class of the matrix 
 $D_\pi=\scriptsize\Big(\begin{array}{cc} 0&1\\-1&0\end{array}\Big)\in{\rm SL(2,\R)}$.
 \end{definition} 
 It is obvious that 
 \[\varphi_t(\T(x))=\T(\varphi_{-t}(x))
 \quad \mbox{for}\quad x\in X\quad\mbox{and}\quad t\in\R.\] 
 
Recall the number $\eps_*$ from Lemma \ref{per-coinc}.
\begin{definition}[Orbit pair/Partner orbit]\label{partnerdf1}
Let $\eps\in \,]0,\eps_*[$ be given. 
Two
given $T$-periodic orbit $c$ and $T'$-periodic orbit $c'$ of the flow
$(\varphi_t)_{t\in\R}$ are called an {\em $\eps$-orbit pair}
if 
there are $L\geq 2, L\in \Z$ and two decompositions  of $[0,T]$ and $[0,T']$: $0=t_0<\cdots<t_L=T$
and $0=t_0'<\cdots<t_L'=T'$, 
and a permutation $\sigma:\{0,1,\dots,L-1\}\rightarrow 
\{0,1,\dots, L-1\}$ such that 
for each $j\in\{0,\dots,L-1\}$, 
either
\begin{equation*}
d_X(\varphi_{t+t_j}(x), \varphi_{t+t_{\sigma(j)}'}(x'))<\eps
\quad \mbox{for all}\quad t\in [0,t_{j+1}-t_j]
\end{equation*}
or
\begin{equation*}
d_X\Big(\varphi_{t+t_j}(x), \varphi_{t-t_{\sigma(j)+1}'}(\T(x'))\big)<\eps
\quad \mbox{for all}\quad t\in [0,t_{j+1}-t_j]
\end{equation*}
holds for some $x\in c$ and $x'\in c'$.
Then $c'$ is called an {\em $\eps$-partner orbit} of $c$ and vice versa. 
\end{definition}
Roughly speaking, two periodic orbits are an $\eps$-orbit pair if they are $\eps$-close to each other in configuration space, not for the whole time, since otherwise they would be identical due to Lemma \ref{per-coinc}, but they decompose to the same number of parts and any part of one orbit is $\eps$-close to some part of the other. 
The above definition is modified from \cite{bieder}. In this paper we will use the following equivalent definition.
\begin{definition}[Orbit pair/Partner orbit]
Let $\eps\in\,]0,\eps_*[$ be given. Two
given $T$-periodic orbit $c$ and $T'$-periodic orbit $c'$ of the flow $(\varphi_t)_{t\in\R}$ are called an {\em $\eps$-orbit pair} if there are a number $L\geq 2, L\in\Z$ and  a permutation $P:\{1,\dots,L\}\rightarrow\{1,\dots,L\}$ satisfying the following conditions.
\begin{enumerate}
         \item[(a)] There are $x_1,\dots, x_n\in c,\ x_1',\dots,x_L'\in c'$ such that
         \begin{eqnarray*} 
          \varphi_{T_j}(x_j)
          &=&\begin{cases} x_{j+1} & \quad\mbox{if}\quad j\ne L\\
                         x_ 1 &\quad  \mbox{if}\quad j= L
                       \end{cases}\hskip1,1cm \mbox{for}\quad T_j>0\quad  \mbox{and} \quad  T=T_1+\cdots+T_L,
                             \\
         \varphi_{T'_{P(j)}}(x_j') &= & 
          \begin{cases} x_{P(j)+1}' &\mbox{if}\quad P(j)\ne L\\
          x_1' &\mbox{if}\quad P(j)=L
          \end{cases}\quad \mbox{for}\quad  T_j'>0\quad \mbox{and} \quad  T'=T_1'+\cdots+T_L';
          \end{eqnarray*}
    \item[(b)] For $j=1,\dots,L$,
    \[\mbox{either}\quad 
  d_X(\varphi_{t}(x_j'),\varphi_{t}(x_{P(j)}))<\eps\quad
  \mbox{or}\quad
   d_X(\varphi_{t-T_{P(j)}'}(\T (x_j')),\varphi_{t}(x_{P(j)}))<\eps \]
holds for all $t\in \big[0,\max\{T_{P(j)},T_{P(j)}'\}\big]$.
   \end{enumerate}
   Then the orbit $c'$ is called an {\em $\eps$-partner orbit}
 of the orbit $c$ and vice versa.
 \end{definition}
 We shall often skip the parameter $\eps$  and  call an {$\eps$-orbit pair}  an {\em orbit pair} and
 call an {\em $\eps$-partner orbit} a {\em partner orbit} or a {\em partner}. 
 It is clear that any periodic orbit always has 2 trivial partner orbits, namely itself and its time reversal orbit. We do not count them as partner orbits.

\begin{figure}[ht]
	\begin{center}
		\begin{minipage}[ht]{0.9\linewidth}
			\centering
			\includegraphics[angle=0,width=1\linewidth]{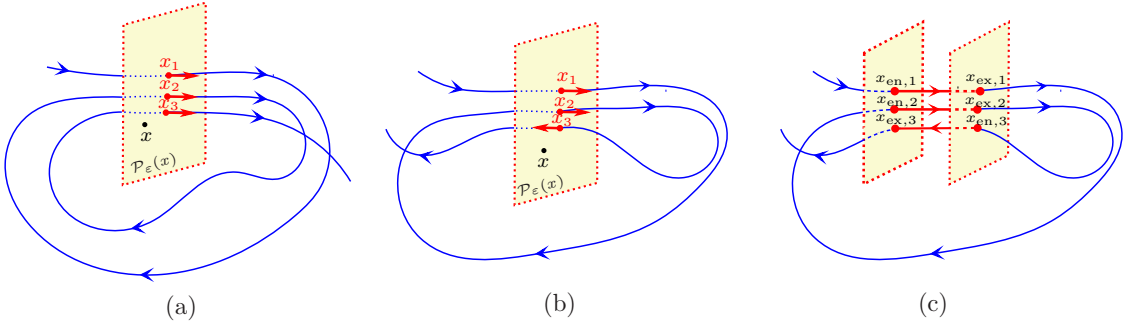}
		\end{minipage}
	\end{center}
	\caption{(a)\,\&\,(b) Example of parallel and anti-parallel encounters; (c) Entrance and exit ports}\label{encounter}
\end{figure}

\begin{definition}[Encounter] Let $\eps\in\,]0,\frac{\eps_*}4[$ and $L\in \Z, L\geq 2$ be given.
 We say that a periodic orbit $c$ has an {\em$(L,\eps)$-encounter} if
there are $x\in X$, $x_1,\dots, x_L\in c$ such that for each $j\in\{1,\dots,L\}$,
\[\mbox{either}\quad x_j\in \P_\eps(x)\quad \mbox{or}\quad \T(x_j)\in \P_\eps(x).\]
If either $x_j\in\P_\eps(x)$ holds for all $i=1,\dots,L$ or $\T(x_j)\in\P_\eps(x)$ holds for all $j=1,\dots,L$  then
the encounter is called {\em parallel encounter}; otherwise it is called {\em antiparallel encounter}. We call the points 
$x_1,\dots, x_L$  {\em piercing points} (see Figure \ref{encounter}\,(a)\,$\&$\,(b) below). 
\end{definition}
The parameter $\eps$ will be often dropped if the radius of the Poincar\'e section is not important. 

\subsubsection*{Encounter duration} 
Given an $(L,\eps)$-encounter with piercing points $x_1,\dots,x_L$. Owing to Lemma \ref{lm1}, 
we can assume that $x_i\in\P_\eps(x_1)$ for $i=2,\dots,L$. 
Note that either $d_X(x_j,x_i)<4\eps<\eps_*$
or $d_X(\T(x_j), x_i)<4\eps<\eps_*$ for any $i,j\in\{1,\dots,L\}$. By the continuity of the flow $(\varphi_t)_{t\in \R}$,
there are $L$ orbit stretches of length $t_{\rm enc}$ through $x_1,\dots,x_L$ which remain $4\eps$-close to each other
(up to time reversal),
and we call these stretches  the {\em encounter region}, 
and $t_{\rm enc}$ is called the {\em encounter duration}. We are going to evaluate $t_{\rm enc}$. First, we consider a 2-parallel encounter. Let $c$ be a $T$-periodic orbit with a $2$-parallel encounter. Assume that $x,y\in c$ such that 
$y\in\P_\eps(x)$ for $y=(u,s)_x$. Then 
\[d_X(x,y)=d_X(\Gamma g,\Gamma gc_ub_s)\leq d_\G(e,c_ub_s)\leq |u|+|s|<2\eps,\]
 where
$g\in\G, \Gamma g=x$.  
Using $b_sa_t=a_tb_{se^{-t}}$ and $c_u a_t=a_tc_{ue^{t}}$, we deduce that 
\[\varphi_t(y)=\Gamma (ga_t)(c_{ue^t}b_{se^{-t}})
\quad\mbox{for all}\quad t\in\R.\]
Then
$\varphi_t(y)\in\P_\eps(\varphi_t(x))$ if and only if $|u|e^t <\eps$ and $|s|e^{-t}<\eps$. 
Note that $us\ne 0$ since otherwise $y\in W_X^s(x)$ or $y\in W_X^u(x)$ which contradicts Lemma \ref{nocom}. 
So, $\varphi_t(y)\in \P_\eps(\varphi_t(x))$
for  $-\ln\Big(\frac{\eps}{|s|}\Big)<t<\ln\Big(\frac{\eps}{|u|}\Big)$.
  Denote $t_s=\ln\Big(\frac{\eps}{|s|}\Big)$ and
$t_u=\ln\Big(\frac{\eps}{|u|}\Big)$. Then for   $t\in\,]-t_s,t_u[$, $\varphi_t(y)\in \P_\eps(\varphi_t(x)), \varphi_t(y)=(ue^t,se^{-t})_{\varphi_t(x)}$ and hence $d_X(\varphi_t(x),\varphi_t(y))<2\eps$.  The encounter duration is thus given by 
\[t_{\rm enc}=t_s+t_u=\ln\Big(\frac{\eps}{|s|}\Big)+\ln\Big(\frac{\eps}{|u|}\Big)=\ln\Big(\frac{\eps^2}{|us|}\Big).\]
We see that $t_s$ is the duration the flow can go backward and 
$t_u$ is the duration the flow can go toward before leaving the encounter region. Next, for an $(L,\eps)$-parallel encounter with piercing points $x_j\in\P_\eps(x_1)$,  $x_j=(u_j,s_j)_{x_1}$
for $j=2,\dots,L$, we define  
\begin{equation}\label{tstu} 
t_s=\min_{2\leq j\leq L}\left\{\ln\Big(\frac{\eps}{|u_j|}\Big)\right\}\quad\mbox{and}\quad t_u=\min_{2\leq j\leq L}\left\{\ln\Big(\frac{\eps}{|s_j|}\Big)\right\}.
\end{equation} 
The encounter duration is given by
\begin{equation}\label{tenc}
t_{\rm enc}=t_s+t_u=\ln\Big(\frac{\eps^2}{us}\Big),
\end{equation} 
where $u=\max_{2\leq j\leq L}\{|u_j|\},\, s=\max_{2\leq j\leq L}\{|s_j|\}$.
For antiparallel encounters, the argument is similar and we also have \eqref{tenc} for the encounter duration; see \cite{mueller2005} for similar results.

\begin{definition}[Entrance port/Exit port] Given  an  $(L,\eps)$-encounter with  piercing points $x_1,\dots,x_L$.  
For $j=1,\dots,L$, we define the {\em entrance port}  and 
the {\em exit port} of the $j$th orbit stretch by
\[x_{{\rm en},\,j}=\begin{cases}
\varphi_{-t_s}(x_j) \quad &\mbox{if}\quad\quad\ \, x_j\in\P_\eps(x_1) \\
\varphi_{-t_u}(x_j) \quad &\mbox{if}\quad \T(x_j)\in\P_\eps(x_1) 
\end{cases}\]
and
\[x_{{\rm ex},\,j}=\begin{cases}
\varphi_{t_u}(x_j) \quad &\mbox{if}\quad\quad\ \,
x_j\in\P_\eps(x_1) \\
\varphi_{t_s}(x_j) \quad &\mbox{if}\quad \T(x_j)\in\P_\eps(x_1), 
\end{cases}\]
respectively; recall $t_s$ and $t_u$ in \eqref{tstu}. 
 \end{definition}
 We see that the $j$th stretch
 enters the encounter region through the entrance port
 $x_{{\rm en},\,j}$ and leaves it through the exit port
 $x_{{\rm ex},\,j}$ (see Figure \ref{encounter}\,(c)).

\begin{example}\label{edex}\rm  Assume that a periodic orbit of the geodesic flow on $T^1(\Gamma\backslash\H^2)$ crosses itself in configuration space at an angle $\theta$ such that
$|\phi| <\min\{1/6,\eps_*/9\}$ for $\phi=\pi-\theta$; see Figure \ref{2encounter}. Then it has a unique partner orbit (see Theorem 3.15 in Part I \cite{HK}). Denote  $\eps=\frac32\,\sin(\phi/2)<\frac{\eps_*}{12}=\varrho$. Recall
$x=\Gamma g, y=\Gamma h, y'=\T(y)=\Gamma h'$.

\smallskip
\noindent
(i) The original orbit has a 2-antiparallel encounter. Indeed,
by the proof of Theorem  3.5  in Part I, 
$x=\Gamma g=\Gamma h' d_\theta=\Gamma h' a_{-\tau}c_{-u}b_{-s}$, 
where $d_\theta\in\PSL(2,\R)$ is the equivalence class of $D_\theta=\scriptsize\Big(\begin{array}{cc}\cos(\theta/2)&\sin(\theta/2)\\
-\sin(\theta/2)&\cos(\theta/2)\end{array}\Big)\in\SL(2,\R), 
\tau=-2\ln(\cos(\phi/2)),\quad s=\tan(\phi/2),\quad u=-\sin(\phi/2)\cos(\phi/2)$. 
 This means that
$x\in \P_{\varrho}(\tilde y)$ for $x=(-u,-s)_{\tilde y}$ with $\tilde y=\varphi_{-\tau}(y')$. 
Consider the orbit $c$ of $x'=\T(x)$. 
It follows from 
$x',\tilde y\in c$ and $\T(x')=x\in\P_\varrho(\tilde y)$ 
that the orbit $c$ has a 2-antiparallel encounter. 

\smallskip 
\noindent
(ii) The partner orbit is a $6\eps$-partner. To see this, put $x_1=x, x_2=y, x_1'=w, x_2'=\varphi_{T_2}(w),
T_1'=T_2, T_2'=T'-T_1'$. 
We have $\varphi_{T_1}(x_1)=x_2,\varphi_{T_2}(x_2)=x_1$
as well as $\varphi_{T_1'}(x_1')=x_2', \varphi_{T_2'}(x_2')=x_1'$.
Furthermore, it was shown in Part I that 
\[
d_X(\varphi_{t}(x_{P(1)}),\varphi_{t-T_{P(1)}'}(\T x_1'))=
d_X(\varphi_{t}(x_2),\varphi_t(\T x_2'))< 6\eps
\quad \mbox{for}\quad t\in[0,T_2]\]
and \[d_X(\varphi_t(x_{P(2)}),\varphi_t(x_2'))=
d_X(\varphi_t(x_1),\varphi_t(x_2'))<6\eps
\quad \mbox{for}\quad t\in [0,T_{P(2)}],\]
here $P=\scriptsize\Big(\begin{array}{cc}1&2\\2&1 \end{array}\Big)$.
Thus, the partner orbit is a $6\eps$-partner orbit.

\smallskip
\noindent 
(iii) 
The encounter duration 
\[t_{\rm enc}=\ln\Big(\frac{\varrho^2}{\sin^2(\phi/2)}\Big) .\]
For more details, see Subsection 3.5 in Part I \cite{HK}. 
 {\hfill$\diamondsuit$}
\end{example}

In the remaining part, encounters mean parallel encounters. 
\subsection{Number of partner orbits}\label{combisec}
Let $c$ be a given periodic orbit with an $L$-encounter ($L\geq 2$). The orbit connects the $j$th entrance port and the $j$th exit port, $j=1,\dots,L$. This can be described by the identical permutation 
$P_{\rm ori}=\Big(\scriptsize\begin{array}{cccc} 1&2&\dots&L\\1&2&\dots&L\end{array}\Big)=e$.
We can connect the entrance ports and the exit ports by different ways to get different partner orbits. 
The way  which entrance port is connected to which exit port can be expressed by a permutation $P\in S_L$. However, not all the permutations in $S_L$ give connected partner orbits.  
A permutation $P$ illustrated a partner orbit has to satisfy
the condition  that $P_{\rm loop}P$ is a single cycle, where 
$P_{\rm loop}=\Big(\scriptsize\begin{array}{ccccc} 1&2&\dots&L-1&L\\2&3&\dots&L&1\end{array}\Big)$
is the {\em orbit loops permutation}, because  it is a periodic orbit and hence returns to the first entrance port only after traversing all others.  Recall that a permutation 
$P\in S_L$ is called a single cycle if $P$ cannot be written as a product of shorter cycles; equivalently, $P^k(j)\ne j$ for all $j\in \{1,\dots, L\}$ and $k\in \{1,\dots,L-1\}$. 
For more details, see \cite{mueller2005}. 
Note that we do not demand the permutation $P$ to be a single cycle as in \cite{mueller2005}.
\begin{lemma}
The number of permutations $P$ in $S_L\setminus \{e\}$ such that $P_{\rm loop} P$ are single cycles is $(L-1)!-1$.
\end{lemma}
\noindent
 {\bf Proof.} It is well-known that the number of single cycles in $S_L$ is $(L-1)!$.  
 For every single cycle $Q$, we write $Q=P_{\rm loop}(P^{-1}_{\rm loop}Q).$ 
 Then the permutation $P=P^{-1}_{\rm loop}Q$ satisfies the condition that $P_{\rm loop}P$ is a single cycle.
 Note that $P_{\rm loop}$ is a single cycle and the identity permutation $e$ also satisfies 
 the condition that $P_{\rm loop}e$ is a single cycle. This completes the proof. 
{\hfill$\Box$}
\begin{example}\rm
\noindent 
(i) For $L=2$, there are $(2-1)!-1=0$ partner orbits. 

\medskip
\noindent 
(ii) For $L=3$, there is $(3-1)!-1=1$ partner orbit
which is illustrated by the permutation $\Big(\scriptsize\begin{array}{ccc} 1&2&3\\2&3&1\end{array}\Big)$;
see Figure \ref{3encounter}\,(a).

\begin{figure}[ht]
	\begin{center}
		\begin{minipage}[ht]{0.9\linewidth}
			\centering
			\includegraphics[angle=0,width=1\linewidth]{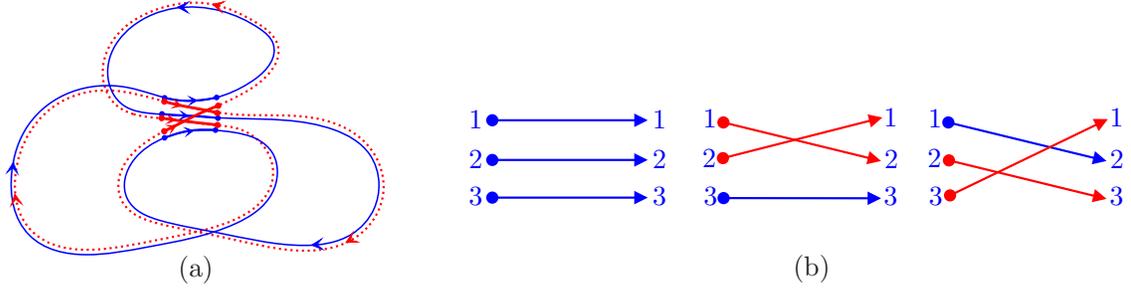}
		\end{minipage}
	\end{center}
	\caption{(a) Partner orbit of periodic orbit with 3-parallel encounter; (b) Interchange entrance and exit ports to creat the partner orbit}\label{3encounter}
\end{figure}

\medskip
\noindent
(iii) For $L=4$, there are $(4-1)!=6$ single cycles in $S_4$, namely:
\[(\begin{array}{cccc}1& 2 & 3& 4\end{array}),\ (\begin{array}{cccc}1& 2 & 4& 3\end{array}),\ (\begin{array}{cccc}1& 3 & 2& 4\end{array}),\ (\begin{array}{cccc}1& 3 & 4& 2\end{array}),\ (\begin{array}{cccc}1& 4 & 3& 2\end{array}),\ (\begin{array}{cccc}1& 4 & 2& 3\end{array}).\]
The first one corresponds to the original orbit and 
therefore there are 5 partner orbits given by the following permutations:
\[\bigg(\begin{array}{cccc} 1&2&3&4\\ 1&3&4&2 \end{array}\bigg),
\ \bigg(\begin{array}{cccc} 1&2&3&4\\ 2&3&1&4 \end{array}\bigg),
\ \bigg(\begin{array}{cccc} 1&2&3&4\\ 2&4&3&1\end{array}\bigg),
\ \bigg(\begin{array}{cccc} 1&2&3&4\\3&4&1&2 \end{array}\bigg), \ 
\bigg(\begin{array}{cccc} 1&2&3&4\\ 3&2&4&1 \end{array}\bigg).\]
{\hfill$\diamondsuit$}
\end{example}

\subsection{3-encounters}\label{3secc}
In this subsection we will construct a partner orbit of a given orbit with a $3$-encounter. 
Denote $P=\Big(\scriptsize\begin{array}{ccc} 1&2&3\\ 2&3&1\end{array}\Big)\in S_3$. 
\begin{theorem}\label{3-encounter}
Let $\eps\in\,]0,\frac{\eps_*}{11}[$ and let $c$ be a $T$-periodic orbit involving a $(3,\frac{\eps}3)$-encounter with the following property:
\begin{enumerate}
\item[(i)] there are $x_j\in c$, $x_j\in\P_{\frac{\eps}3}(x)$ with $x_j=(u_j,s_j)_{x}$ for some $x\in X$, $j=1,2,3$;
\item[(ii)] $|u_j-u_i|>\frac{6}{5}(e^{-T_j}+e^{-T_i})$
 and $|s_j-s_i|>\frac{24}{3^3}\eps^3+e^{-T_{j-1}}+e^{-T_{i-1}}$ for $j,i=1,2,3$ and $j\ne i$, where
$T_1,T_2,T_3>0$ are such that $T=T_1+T_2+T_3$,
$\varphi_{T_1}(x_1)=x_2,\varphi_{T_2}(x_2)=x_3,$ and $\varphi_{T_3}(x_3)=x_1$.
\end{enumerate}
Then the orbit $c$ has a $11\eps$-partner $ c'$ whose period $T'$ satisfies
\begin{eqnarray}\label{actionL=3}
\Big|\frac{T'-T}2-\Delta S_3\Big|<76\eps^4+22\eps^2 e^{-T_1}+7\eps^2 e^{-T_2}+7\eps^2e^{-T_3},
\end{eqnarray}
where \[\Delta S_3=\ln(1+(u_3-u_1)( s_3- s_2))+\ln(1+( u_2-u_1)( s_2-s_1)).\]
Furthermore, the partner orbit also has a $(3,\eps)$-encounter.
More precisely, there are $v_1,v_2,v_3\in  c'$ and $T'_1,\,T'_2,\,T'_3>0$ such that
\begin{enumerate}
\item[(a)] $\varphi_{T'_2}(v_1)=v_3,\, \varphi_{T'_3}(v_2)=v_1,\, \varphi_{T'_1}(v_3)=v_2$, and $T'=T_1'+T_2'+T_3'$; 
\item[(b)]  $v_j\in\P_\eps(x)$ with $v_j=(u'_j,s'_j)_x$ satisfying
\begin{subequations}\label{encounter3}
\begin{alignat}{2}
\label{uj'3}
|u'_j-u_{P(j)}|&\,<23\eps^3+ 6\eps (e^{-T_1}+e^{-T_2}+e^{-T_3}),
\\
\label{sj'3}
|s'_j-s_j|&\,<\,23\eps^3+ 6\eps(e^{-T_1}+e^{-T_2}+e^{-T_3}); 
\end{alignat}
\end{subequations}
\item[(c)] $ |T'_j-T_j|<22\eps^2$;
\item[(d)] $d_X(\varphi_t(v_j),\varphi_t(x_{P(j)}))<10\eps$\quad for all\quad $t\in [0, \max\{T_{P(j)},T'_{P(j)}\}]$.
\end{enumerate}
\end{theorem}
\noindent{\bf Proof.}
First we use Lemma \ref{lm1} to write
$\tilde x_j:=\varphi_{\tilde\tau_j}(x_j)\in\P_{\eps}(x_1)$ with $\tilde x_j=(\tilde u_j,\tilde s_j)_{x_1}$ and
\begin{eqnarray}\label{tilde-}
|\tilde u_j-(u_j-u_1)|<\frac{1}{3}\,\eps^3,\quad  
|\tilde s_j-(s_j-s_1)|<\frac{1}{3}\,\eps^3,
\quad |\tilde u_j\tilde s_j|<\frac{5}{9}\,\eps^2,\quad|\tilde \tau_j|<\frac{8}9\,\eps^2.
\end{eqnarray} 
Denoting
\begin{eqnarray}\label{tildeT}
\tilde T_1= T_1+\tilde \tau_2,\quad \tilde T_2=T_2-\tilde \tau_2+\tilde\tau_3,\,\,\, \mbox{and }\,\, \tilde T_3=T_{3}-\tilde \tau_3,
\end{eqnarray}   we have
\begin{equation}\label{tildex}
\varphi_{\tilde T_1}(x_1)=\tilde x_{2},\quad \varphi_{\tilde T_2}(\tilde x_2)=\tilde x_3,\quad \varphi_{\tilde T_3}(\tilde x_3)=x_1,
\end{equation}
and $T=\tilde T_1+\tilde T_2+\tilde T_3$
as well as
\begin{equation}\label{T1T2-}
|T_1-\tilde T_1|<\frac{8}{9}\,\eps^2,\quad 
|T_2-\tilde T_2|<2\eps^2,\quad \mbox{and}\quad 
|T_3-\tilde T_3|<\frac{8}{9}\,\eps^2.
\end{equation}
Figure \ref{3encounter} illustrates steps of interchange ports in a 3-encounter to create a partner orbit.

\medskip 

\underline{Step 1:} 
Write $x_j=\Gamma g_j$ and $\tilde x_j=\Gamma\tilde g_j$ for $g_j,\tilde g_j\in\G$, $j=1,2,3$.
By the above setting, we have  $\tilde x_2=\varphi_{\tilde T_1}(x_1)=(\tilde u_2,\tilde s_2)_{x_1}\in\P_{\eps}(x_1)$. 
Apply the Anosov closing lemma I (Theorem \ref{anosov1}) to obtain
$y_2=\Gamma h_2=\Gamma g_1c_{\sigma_2}b_{\eta_2}\in \P_{2\eps}(x_1)$ and $\hat T_1\in\R$ so that $\varphi_{\hat T_1}(y_2)=y_2$,
\begin{equation}\label{y2x1}
d_X(\varphi_t(x_1),\varphi_t(y_2))<4\eps\quad\mbox{for all}\quad t\in[0,\tilde T_1],
\end{equation}
\begin{equation}\label{hatT1-}
 \Big|\frac{\hat T_1-\tilde T_1}{2}-\ln(1+\tilde u_2\tilde s_2)\Big|< 5|\tilde u_2\tilde s_2|e^{-\tilde T_1},
\end{equation}
and 
\begin{eqnarray}\label{sigma2eta2}
|\sigma_2|< 2\eps e^{-\tilde T_1},\quad
|\eta_2-\tilde s_2|<2\eps^3+2 \eps e^{-\tilde T_1}.
\end{eqnarray}
Then  
\begin{equation}\label{hatT1-tildeT1,3}
|\hat T_1-\tilde T_1|<5|\tilde u_2\tilde s_2|<3\eps^2 
\end{equation}
yields
\begin{eqnarray}\label{hatT1-T1,3}
|\hat T_1-T_1|
\leq
|\hat T_1-\tilde T_1|+|\tilde T_1- T_1|
<4\eps^2
\end{eqnarray}
due to \eqref{T1T2-}. 
Furthermore,
\begin{eqnarray*}\label{s1+eta2,3}
|\eta_2+s_1|
        \leq |\eta_2-\tilde s_2|+|\tilde s_2-(s_2-s_1)|+|s_2|
                \leq 2\eps^3+2\eps e^{-\tilde T_1}+\frac{1}{3}\,\eps^3+\frac{1}{3}\,\eps<\eps   
\end{eqnarray*}
yields $|\eta_2|<\frac{4}{3}\,\eps|$.
On the other hand, note that $\varphi_{\tilde T_2+\tilde T_3}(\tilde x_2)=x_1
=\Gamma  \tilde g_2 b_{-\tilde s_2}c_{-\tilde u_2}\in\P'_{\eps}(\tilde x_2)$ 
and $\varphi_{\tilde T_2+\tilde T_3}(\tilde x_2)=(-\tilde u_2,-\tilde s_2)_{\tilde x_2}$. Then, by the Anosov closing lemma II (Theorem \ref{anosov2}),
 there are $y_1=\Gamma h_1=\Gamma  \tilde g_2 b_{\eta_1}c_{\sigma_1}\in \P'_{2\eps}(x_2)$
and $\hat T_{2,3}\in\R$ 
so that $\varphi_{\hat T_{2,3}}(y_1)=y_1$,
\begin{equation}\label{y1x2}
d_X(\varphi_t(y_1),\varphi_t(\tilde x_2))<4\eps\quad \mbox{for all}\quad t\in [0,\tilde T_2+\tilde T_3],
\end{equation}
\begin{equation}\label{hatT2,3-}
 \Big|\frac{\hat T_{2,3}-({\tilde T_2+\tilde T_3})}{2} \Big|
< 4|\tilde u_2\tilde s_2|e^{-\tilde T_2-\tilde T_3},
\end{equation}
and
\begin{eqnarray}\label{sigma1eta1L=3}
          \quad |\sigma_1|<2\eps e^{-\tilde T_2-\tilde T_3},\quad
 |\eta_1+\tilde s_2|<2\eps e^{-\tilde T_2-\tilde T_3}.
        \end{eqnarray}
Then
\begin{equation}\label{hatT23-}
\big|\hat T_{2,3}-(\tilde T_2+\tilde T_3)\big|
<\frac{40}{9}\,\eps^2 e^{-\tilde T_2-\tilde T_3}<\eps^2
\end{equation}
and 
$|\eta_1|<\frac{4}{3}\,\eps$.

\medskip 

 \underline{Step 2:} {\em Construction of the partner orbit}.
We are going to `connect' the orbits of $y_1$ and $y_2$ to get a new periodic orbit.
We need to check the assumption of the connecting lemma (Lemma \ref{connect0}). 
 Define $y_3=\Gamma h_3=\varphi_{\tilde T_2}(y_1)$ and recall  
 $y_1=\Gamma \tilde g_2 b_{\eta_1}c_{\sigma_1},\, \Gamma\tilde g_3=\Gamma g_1 c_{\tilde u_3}b_{\tilde s_3}$,
  and $\Gamma h_2=\Gamma g_1c_{\tilde u_2}b_{\tilde s_2}$. This implies that
\begin{eqnarray*}
y_3 &=&\Gamma  \tilde g_2 b_{\eta_1}c_{\sigma_1}a_{\tilde T_2}
=\Gamma  \tilde g_2a_{\tilde T_2}b_{\eta_1e^{-\tilde T_2}}c_{\sigma_1 e^{\tilde T_2}}
=\Gamma  \tilde g_3 b_{\eta_1 e^{-\tilde T_2}}c_{\sigma_1e^{\tilde T_2}}
\\
&=&\Gamma g_1c_{\tilde u_3}b_{\tilde s_3+\eta_1 e^{-\tilde T_2}}c_{\sigma_1e^{\tilde T_2}}
=\Gamma h_2b_{-\eta_2}c_{\tilde u_3-\sigma_2}b_{\tilde s_3+\eta_1 e^{-\tilde T_2}}c_{\sigma_1e^{\tilde T_2}}.
\end{eqnarray*}
Applying Lemma \ref{bcbcbc}, we write 
$y_3=\Gamma h_2 c_{\check u_3}b_{\check s_3}a_{\check \tau_3}$
for
\begin{eqnarray*}\label{checku3}
\check u_3&=&\tilde u_3-\sigma_2+\sigma_1e^{\tilde T_2}
+\frac{1}{1+\check\rho_3}\,\big((\tilde u_3-\sigma_2)\sigma_1e^{\tilde T_2}(\tilde s_3+\eta_1e^{-\tilde T_2})
-(\tilde u_3-\sigma_2+\sigma_1e^{\tilde T_2})\check\rho_3\big),\\ \label{checks3}
 \check s_3&=&-\ \eta_2+\tilde s_3+\eta_1e^{-\tilde T_2}
 +\check\rho_3(-\eta_2+\tilde s_3+\eta_1e^{-\tilde T_2})
 -(\tilde u_3-\sigma_2)\eta_2(\tilde s_3+\eta_1e^{-\tilde T_2})(1+\check\rho_3),\\ \nonumber
\check \tau_3&=& 2\ln(1+\check\rho_3),
\end{eqnarray*}
where
\[\check\rho_3=\sigma_1e^{\tilde T_2}(-\eta_2+\tilde s_3+\eta_1e^{-\tilde T_2})
-(\tilde u_3-\sigma_2)\eta_2(1+\sigma_1e^{\tilde T_2}(\tilde s_3+\eta_1e^{-\tilde T_2})).
\]
From $|\eta_1|,|\eta_2|<\frac{4\eps}{3} $, \eqref{sigma2eta2}, and \eqref{sigma1eta1L=3}, we thus have
$|\check \rho_3|<3\eps^2$ and hence
\begin{eqnarray}
\label{checktau3}
|\check\tau_3| <12\eps^2, \quad |\check u_3-\tilde u_3|<
 7\eps^3+ 2\eps e^{-\tilde T_1}+2\eps e^{-\tilde T_3}, \quad 
|\check s_3-(-\eta_2+\tilde s_3)|<16\eps^3+2\eps e^{-\tilde T_2}.
\end{eqnarray}
Therefore 
\begin{eqnarray}\label{checku3-}
|\check u_3-(u_3-u_1)|\leq 
 |\tilde u_3-(u_3-u_1)|+|\check u_3-\tilde u_3|
 \leq 8\eps^3+ 2\eps e^{-\tilde T_1}+2\eps e^{-\tilde T_3}
 \end{eqnarray}
 as well  as
 \begin{eqnarray}\label{checks3-}
|\check s_3-(s_3-s_2)|
&\leq& |\check s_3-(-\eta_2+\tilde s_3)|
+|\tilde s_2-\eta_2|
+|(s_2-s_1)-\tilde s_2|
+|\tilde s_3-(s_3-s_1)|
\notag\\ 
&<& 19\eps^3+2\eps e^{-\tilde T_1}+2\eps e^{-\tilde T_2},
\end{eqnarray}
using
\eqref{tilde-} and \eqref{sigma2eta2}, so that $|\check u_3|<\eps,|\check s_3|<\eps$, 
and hence
\begin{equation}\label{checky3,3}
 \check y_3:=\Gamma \check h_3=\Gamma h_3a_{-\check \tau_3}=\Gamma h_2c_{\check u_3}b_{\check s_3}\in \P_{\eps}(y_2).
\end{equation} 
Recall that $y_2=\Gamma h_2$ is $\hat T_1$-periodic and since $y_3=\Gamma h_3$ is $\hat T_{2,3}$-periodic, so is $\check y_3=\varphi_{-\check\tau_3}(y_3)$. Apply Lemma \ref{connect0} to obtain
$z_3=\Gamma h_2 c_{\check u_3 e^{-\hat T_1}+\sigma} b_\eta\in\P_{2\eps}(y_2)$
and $T'\in \R$ such that
$\varphi_{T'}(z_3)=z_3$,
\begin{eqnarray}\label{z3y2}
d_X(\varphi_t(z_3),\varphi_t( y_2)) &<& 5\eps\quad\mbox{for all}\quad t\in[0,\hat T_1],
\\ \label{z3checky3}
d_X(\varphi_{t+\hat T_1}(z_3),\varphi_t(\check y_3)) &<& 5\eps\quad\mbox{for all}\quad t\in[0,\hat T_{2,3}],
\end{eqnarray}
\begin{eqnarray}\label{T'-cc}
 \Big|\frac{T'-(\hat T_1+\hat T_{2,3})}{2}-\ln(1+\check u_3\check s_3)\Big|
< 5|\check u_3\check s_3|(e^{-\hat T_1}+e^{-\hat T_{2,3}}),
\end{eqnarray}
and
\begin{equation}\label{sigmaeta}
|\sigma|<2\eps e^{-\hat T_1-\hat T_{2,3}},\quad 
|\eta-\check s_3|< 2\eps^3+2\eps e^{-\hat T_1-\hat T_{2,3}}
< 2\eps^3+\eps e^{-T_2}.
\end{equation}
 Using \eqref{hatT1-T1,3} and \eqref{hatT2,3-}, 
 we rewrite \eqref{T'-cc} as 
\begin{eqnarray} \label{T'-(hatT1+hatT2,3)}
 \Big|\frac{T'-(\hat T_1+\hat T_{2,3})}{2}-\ln(1+\check u_3\check s_3)\Big|
 <6|\check u_3\check s_3|(e^{-\tilde T_1}+e^{-\tilde T_2-\tilde T_3}).
\end{eqnarray}

\underline{Step 3:} {\em Proof of \eqref{actionL=3}}. 
We are in a position to derive an estimate for the action difference.
 For, it follows from \eqref{T'-(hatT1+hatT2,3)}, \eqref{hatT1-tildeT1,3}, and \eqref{hatT2,3-} that
\begin{eqnarray}\label{(T'-T),3} \notag 
&&\hskip-1cm\Big|\frac{T'-T}{2}-\Big(\ln(1+\check u_3\check s_3)+\ln(1+\tilde u_2\tilde s_2)\Big)
\Big|\\ \notag
&&\leq\, 6|\check u_3\check s_3|(e^{-\tilde T_1}+e^{-\tilde T_2-\tilde T_3})
+ 5|\tilde u_2\tilde s_2|e^{-\tilde T_1}+4|\tilde u_2\tilde s_2|e^{-\tilde T_2-\tilde T_3}\\
&&\leq\, 11\eps^2e^{-\tilde T_1}+10\eps^2e^{-\tilde T_2-\tilde T_3}.
\end{eqnarray}
Denote 
\[\Delta S_3=\ln(1+(u_3-u_1)( s_3- s_2))+\ln(1+( u_2-u_1)( s_2-s_1))\]
and \[\Delta S_3'= \ln(1+\check u_3\check s_3)+\ln(1+\tilde s_2\tilde s_2).\] 
Using the fact that 
$|\ln(1+x)-\ln(1+y)|<3|x-y|$ for $|x|,|y|<\frac14$, we have 
\begin{eqnarray*}
|\Delta S_3'-\Delta S_3|
&\leq& |\tilde u_2\tilde s_2-(u_2-u_1)(s_2-s_1)|
+3|\check u_3\check s_3-(u_3-u_1)(s_3-s_2)|
\\
&<& 76\eps^2+10 \eps^2 e^{-\tilde T_1} +6\eps^2 e^{-\tilde T_2} +6\eps^2 e^{-\tilde T_3}
\end{eqnarray*}
due to Lemma \ref{lm1} and 
\begin{eqnarray*}
|\check u_3\check s_3-(u_3-u_1)(s_3-s_2)|
&\leq& |\check u_3||\check s_3-(s_3-s_2)|+|\check u_3-(u_3-u_1)||s_3-s_2|
\\
&<& 25\eps^4+\frac{10}3\,\eps^2 e^{-\tilde T_1}+2\eps^2 e^{-\tilde T_2}+2\eps^2e^{-\tilde T_3},
\end{eqnarray*}
by \eqref{checku3-} and \eqref{checks3-}.
Therefore, it follows from \eqref{(T'-T),3} that
\begin{eqnarray*}
\Big|\frac{T'-T}2-\Delta S_3\Big|
&\leq & 76\eps^4+21\eps^2 e^{-\tilde T_1}+\frac{13}2\,\eps^2 e^{-\tilde T_2}+\frac{13}2\,\eps^2 e^{-\tilde T_3}\\
& <& 76\eps^4+22\eps^2 e^{-T_1}+7\eps^2 e^{-T_2}+7\eps^2e^{-T_3},
\end{eqnarray*}
which is \eqref{actionL=3}.

\medskip

\underline{Step 4:} {\em Definition of $v_1,v_2,v_3$ and proof of \eqref{encounter3}.}
In what follows we often use the fact that
\[|\eta_1|<\frac{4}{3}\,\eps,\,|\eta_2|<\frac{4}{3}\,\eps,\, |\sigma_1|<2\eps e^{-\tilde T_2-\tilde T_3},
|\sigma_2|<2\eps e^{-\tilde T_1},\, |\eta|<2\eps,\, |\sigma|<2\eps e^{-\hat T_1-\hat T_{2,3}}.\]
\underline{Step 4.1:} {\em Definition of $v_3$.} Recall that $\Gamma g_1=\Gamma gc_{u_1}b_{s_1},
\Gamma h_2=\Gamma g_1 c_{\sigma_2}b_{\eta_2}, z_3=\Gamma h_2 c_{\check u_3 e^{-\hat T_1}+\sigma}b_\eta$.
Then using  Lemma \ref{bcbcbc}, we can write
\begin{eqnarray*}
z_3=\Gamma h_2 c_{\check u_3 e^{-\hat T_1}+\sigma}b_\eta
=\Gamma g_1 c_{\sigma_2}b_{\eta_2}c_{\check u_3 e^{-\hat T_1}+\sigma}b_\eta
=\Gamma g c_{u_1}b_{s_1}c_{\sigma_2}b_{\eta_2}c_{\check u_3 e^{-\hat T_1}+\sigma}b_\eta
=\Gamma g c_{u'_3}b_{s'_3}a_{\tau'_3}
\end{eqnarray*}
and after a short estimate, we have
\[
|\tau_3'|< \eps^2,\quad 
 |s'_3-(s_1+\eta_2+\eta)|<\eps e^{-T_1},\quad |u_3'-u_1|<4\eps e^{-T_1}.
 \] 
 Therefore, it follows from  \eqref{sigmaeta}, \eqref{checks3-},
 and \eqref{tilde-} that
\begin{eqnarray*}
 |s'_3-s_3|
&\leq& |s'_3-(s_1+\eta_2+\eta)|+|\eta-\check s_3|+|\check s_3-(-\eta_2+\tilde s_3)|+|\tilde s_3-(s_3-s_1)|
\\ \label{s'3a}
&<&23\eps^3+5\eps e^{-T_1}+4\eps e^{-T_2}
\end{eqnarray*}
and hence $|u_3'|<\eps,|s_3'|<\eps$.
Defining $v_3:=\varphi_{-\tau_3'}(z_3)$, we have proven that $v_3=\Gamma g c_{u_3'}b_{s_3'}\in \P_{\eps}(x)$ and $ v_3=(u_3',s_3')$ 
satisfies \eqref{d(n-1)}.
\underline{Step 4.2:} {\em Definition of $v_2$.} Defining 
\begin{equation}\label{z2}
z_2:=\varphi_{\hat T_1}(z_3)
\end{equation}
and using Lemma \ref{bcbcbc}, we write
\begin{eqnarray*}
z_2&=&\Gamma h_2 c_{\check u_3+\sigma e^{\hat T_1}}b_{\eta e^{-\hat T_1}}
=\Gamma  g_1 c_{\sigma_2}b_{\eta_2} c_{\check u_3+\sigma e^{\hat T_1}}b_{\eta e^{-\hat T_1}}\\
&=&\Gamma g c_{u_1}b_{s_1}c_{\sigma_2}b_{\eta_2} c_{\check u_3+\sigma e^{\hat T_1}}b_{\eta e^{-\hat T_1}}
=\Gamma  g c_{u'_2}b_{s'_2}a_{\tau'_2},
\end{eqnarray*}
and after a short calculation, we obtain
 \[|\tau_2'|<5\eps^2,
\quad 
 |s'_2-(s_1+\eta_2)|<2\eps^3+2\eps e^{-\hat T_1},
 \quad 
|u'_2-(u_1+\check u_3)|<\eps^3+3\eps e^{-T_1}+\eps e^{-T_3}.
\]
Together with \eqref{sigma2eta2}, \eqref{tilde-},
and \eqref{checku3-}, 
this yields 
\begin{eqnarray*}
\notag
|s'_2-s_2|
< 5\eps^3+5\eps e^{-T_1}\quad \mbox{as well as}\quad |u'_2-u_3|< 13\eps^3+6\eps e^{-T_1}+4\eps e^{-T_3},
\end{eqnarray*}
and therefore $|u_2'|<\eps,|s_2'|<\eps$.
 Defining $v_2=\varphi_{-\tau'_2}(z_2)$,
  we have proven that $v_2=\Gamma g c_{u'_2}b_{s'_2}\in\P_\eps(x)$ and $v_2=(u'_2,s'_2)_x$ satisfies \eqref{encounter3}.
\underline{Step 4.3:} {\em Definition of $v_1$.}
First, recall that $z_3=\Gamma h_2 c_{\check u_3 e^{-\hat T_1}+\sigma}b_{\eta}$. Then, since 
$y_2=\Gamma h_2$ is a $\hat T_1$-periodic point and
$\Gamma h_3a_{-\check\tau_3}b_{-\check s_3}=\Gamma h_2 c_{\check u_3}$ obtained from \eqref{checky3,3}, we have
  \begin{equation}\label{z3,3}
  z_3=\Gamma h_2 c_{\check u_3 e^{-\hat T_1}+\sigma}b_{\eta}
       =\Gamma h_2 c_{\check \tau_3} a_{-\hat T_1} c_\sigma b_\eta
     =\Gamma h_3a_{-\check \tau_3}b_{-\check u_3}a_{-\hat T_1}c_\sigma b_\eta. 
     \end{equation} 
    Put $\hat T_3=\hat T_{2,3}-\tilde T_2+\check\tau_3,\, 
  \hat T_{1,3}=\hat T_1+\hat T_3$
and define 
\begin{equation}\label{z1,3} 
z_1:=\varphi_{\hat T_3}(z_2).
\end{equation}
Due to $z_2=\varphi_{\hat T_1}(z_3)$ and \eqref{z3,3}, 
it follows that 
\begin{eqnarray*}
z_1=\varphi_{\hat T_{1,3}}(z_3)
=\Gamma h_3a_{\hat T_{2,3}-\tilde T_2}b_{-\check u_3 e^{-\hat T_3}}c_{\sigma e^{\hat T_{1,3}}} b_{\eta e^{-\hat T_{1,3}}}
=\Gamma h_3 a_{-\tilde T_2}b_{-\check u_3 e^{-\hat T_3}}c_{\sigma e^{\hat T_{1,3}}} b_{\eta e^{-\hat T_{1,3}}}; 
\end{eqnarray*} 
recall that $y_3=\Gamma h_3$ is $\hat T_{2,3}$-periodic.
Using $y_3=\Gamma h_3=\varphi_{\tilde T_2}(y_1),\, y_1=\Gamma h_1=\Gamma \tilde g_2 b_{\eta_1}c_{\sigma_1}$ 
and applying Lemma \ref{bcbcbc}, we can write
\begin{eqnarray*}
z_1 &=& 
\Gamma h_1 b_{-\check u_3 e^{-\hat T_3}}c_{\sigma e^{\hat T_{1,3}}} b_{\eta e^{-\hat T_{1,3}}}
=\Gamma g_2 a_{\tilde \tau_2} b_{\eta_1}c_{\sigma_1}b_{-\check s_3 e^{-\hat T_3}}
c_{\sigma e^{\hat T_{1,3}}}b_{\eta e^{-\hat T_{1,3}}}\\
&=&\Gamma gc_{u_2}b_{s_2+\eta_1 e^{\tilde\tau_2}}c_{\sigma_1 e^{-\tilde\tau_2}}
b_{-\check s_3 e^{-\hat T_3+\tilde\tau_2}}c_{\sigma e^{\hat T_{1,3}-\tilde\tau_2}}
b_{\eta e^{-\hat T_{1,3}+\tilde\tau_2}}a_{\tilde\tau_2}\\
&=&\Gamma g c_{u'_1}b_{s'_1}a_{\tau'_1},
\end{eqnarray*}
with $u_1',s_1',\tau_1'$ satisfy
\begin{eqnarray*}
|\tau_1'| < 2\eps^2, \quad  
|u_1'-u_2| 
\leq  \eps^3+3\eps e^{-T_2},
\quad 
|s_1'-(s_2-\eta_1)| \leq  3\eps^3+3\eps e^{-T_3}
\end{eqnarray*}
owing to 
\[\eta_1 e^{\tilde\tau_2}=\eta_1(1+s_1(u_2-u_1))^2=\eta_1+2\eta_1 s_1(u_2-u_1)+\eta_1 s_1^2(u_2-u_1)^2.\]
Together with \eqref{tilde-} and \eqref{sigma1eta1L=3},
this implies 
\begin{eqnarray}\label{s1'a}
|s_1'-s_1|
\leq |s_1'-(s_2-\eta_1)|+|\eta_1+\tilde s_2|+|\tilde s_2-(s_2-s_1)|
\leq 4\eps^3+4\eps e^{-T_3}.
\end{eqnarray}
Defining  
$v_1:=\varphi_{-\tau'_1}(z_1)=\Gamma g c_{u_1'}b_{s_1'}$ leads to 
 $v_1\in\P_\eps(x)$ with $v_1=(u'_1,s'_1)_{x}$ satisfying
\eqref{encounter3} for $j=1$. 

\smallskip

\underline{Step 5:} {\em Proof of (a)$\&$(c).} Recall from \eqref{z2} and \eqref{z1,3} that 
 $z_2=\varphi_{\hat T_1}(z_3)$
 and $z_1=\varphi_{\hat T_3}(z_2)$. 
Letting   
$\hat T_2= T'-(\hat T_1+\hat T_3)$, we have $\varphi_{\hat T_2}(z_2)=z_3$. 
From $v_j=\varphi_{-\tau'_j}(z_j),\, j=1,2,3$, we define 
\begin{equation}\label{Tj'dn}
T_1'=\hat T_1+\tau'_3-\tau'_2,\quad T_2'= \hat T_2-\tau_3'+\tau_1', \quad
T_3'=\hat T_3+\tau'_2-\tau'_1
\end{equation} 
  to obtain   $T'=T_1'+T_2'+T_3'$ and
  \[\varphi_{T'_2}(v_1')=v_3',\quad \varphi_{T'_3}(v_2')=v_1',
  \quad  \varphi_{T_1'}(v_3')=v_2',\]
so  (a) is shown.  Now we show (c). 
It what follows we will use the following result several times:
\begin{equation}\label{th}
|\tau_1'|<2\eps^2,\quad |\tau_2'|< 5\eps^2,\quad
|\tau_3'|<\eps^2,\quad |\check\tau_3|<12\eps^2.
\end{equation}
 First, it follows from \eqref{Tj'dn} that 
\begin{eqnarray*}
|T_1'-T_1|
\leq 
|T_1'-\hat T_1|+|\hat T_1- T_1|
\leq |\tau_3'|+|\tau_2'|+ |\hat T_1-T_1|
\leq 10\eps^2
\end{eqnarray*}
due to \eqref{hatT1-T1,3}.
A short calculation shows that
\[T_2'-T_2=T'-(\hat T_1+\hat T_{2,3})+ \tilde T_2-T_2-\check \tau_3-\tau_3'+\tau_1'.\]
Hence, by \eqref{T1T2-} and \eqref{th}, 
\begin{eqnarray*}
|T_2'-T_2|\leq|T'-(\hat T_1+\hat T_{2,3})|+ |\tilde T_2-T_2|+|\check \tau_3|+|\tau_3'|+|\tau_1'|<22\eps^2;
\end{eqnarray*}
here we have used  
$|T'-(\hat T_1+\hat T_{2,3})|<5\eps^2$ obtained from \eqref{T'-(hatT1+hatT2,3)}.
Finally,
recall that $\hat T_3=\hat T_{2,3}-\tilde T_2+\check{\tau}_3$. Then
it follows from \eqref{Tj'dn} that 
\begin{eqnarray*}
|T_3'-T_3|\leq 
|T_3'-\hat T_3|+|\hat T_3-T_3|
\leq  |\hat T_{2,3}-\tilde T_2-\tilde T_3|+|T_3-\tilde T_3|+|\check\tau_3|+|\tau'_2|+|\tau'_1|
<21\eps^2,
\end{eqnarray*}
using \eqref{hatT23-}, \eqref{T1T2-}, and \eqref{checktau3}. 
In summary, $|T_j'-T_j|<22\eps^2$ for $j=1,2,3$ as was to be shown
in (c). 

\smallskip

\underline{Step 6:} {\em Proof of (d).} 
It what follows we often use the fact that if $z,v\in X$ and
$z=\varphi_\tau(v)$ for some $\tau\in\R$, then
\[d_X(\varphi_t(v),\varphi_t(z))\leq d_\G(a_\tau,e)<|\tau|
\quad \mbox{for all}\quad t\in\R.\]
This applies to                $v_j=\varphi_{-\tau'_j}(z_j), j=1,2,3$. 
\underline{Step 6.1:} For $j=1$. 
For $t\in [0,\min\{\tilde T_1,\hat T_1\}]$,
\begin{eqnarray*}
  d_X(\varphi_{t}(v_3),\varphi_{t}(x_1))
  &\leq& d_X(\varphi_t(v_3),\varphi_t(z_3))
  +d_X(\varphi_t(z_3),\varphi_t(y_2))
         +d_X(\varphi_t(y_2),\varphi_t(x_1))\\
  &\leq& |\tau_3'|+5\eps+4\eps
  < 9\eps+\eps^2,
\end{eqnarray*}
due to \eqref{y2x1} and \eqref{z3y2}. Therefore, since $|T_1-\tilde T_1|<\eps^2,|T'_1-\hat T_1|<6\eps^2$, we have for all $t\in [0,\max\{T_1,T_1'\}]$: 
\begin{eqnarray*}
 d_X(\varphi_{t}(v_3),\varphi_{t}(x_1))
 <
 9\eps+ \eps^2+\sqrt 2\max\{|T-\tilde T_1|, |T_1'-\hat T_1|\}<10\eps,
\end{eqnarray*}
using Lemma \ref{lemT-T'}, and thus (d) is obtained for $j=1$. 
\underline{Step 6.2:} For $j=2$. 
Recall that $z_2=\varphi_{\hat T_1}(z_3),\, y_3=\varphi_{\tilde T_2}(y_2),\ \tilde x_3=\varphi_{\tilde T_2}(\tilde x_2),
\ \check y_3=\varphi_{-\check \tau_3}(y_3)$, and $
\tilde x_3=\varphi_{\tilde \tau_3}(x_3)$. It follows from \eqref{y1x2},
 \eqref{z3checky3}, \eqref{tilde-}, and \eqref{th} that for $t\in [0,\tilde T_3]$, 
\begin{eqnarray*}
d_X(\varphi_t(v_2),\varphi_t(x_3))
&<& d_X(\varphi_t(v_2), \varphi_t(z_2))
     +d_X(\varphi_t(z_2),\varphi_t(\check y_3))
     +d_X(\varphi_t(\check y_3),\varphi_t(y_3))
\\ 
&&  +\,\,d_X(\varphi_t(y_3),\varphi_t(\tilde x_3))
     +
    d_X(\varphi_t(\tilde x_3),\varphi_t(x_3))
     \\
    & < & |\tau'_2|+d_X(\varphi_{t+\hat T_1}(z_3),\varphi_t(\check y_3))
    +|\check\tau_3|+d_X(\varphi_{t+\tilde T_2}(y_1),\varphi_{t+\tilde T_2}(\tilde x_2))
    + |\tilde\tau_3|
    \\
    &<& 9\eps+18\eps^2.
\end{eqnarray*}
Owing to $|T_3-\tilde T_3|<\eps^2$ and $|T_3'-T_3|<22\eps^2$, we apply Lemma \ref{lemT-T'} to obtain (d) for $j=2$. 

\noindent
\underline{Step 6.3:} For $j=3$. The argument is analogous.

\medskip 
\underline{Step 7:} {\em The distinction between the partner orbit and the original orbit}. We skip it and will prove 
it in the next theorem. 
{\hfill$\Box$}\smallskip

\begin{remark}\rm  
(a) According to Step 3 of the proof, the action difference between the orbit pair satisfies
 \begin{eqnarray*}
 \Big|\frac{T'-T}{2}-\Big(\ln(1+\check u_3\check s_3)+\ln(1+\tilde u_2\tilde s_2)
\big)\Big|
\leq 7(|\check u_3\check s_3| +|\tilde u_2\tilde s_2|)(e^{-T_1}+e^{-T_2-T_3}). 
\end{eqnarray*}
(b) The term $76\eps^4$ in \eqref{actionL=3} arises from the coordinate changes
$(\check u_3,\check s_3)\rightarrow (u_3-u_1,s_3-s_2)$
and $(\check u_2, \check s_2)\rightarrow (u_2-u_1, s_2-s_1)$; 
and it cannot be avoided. 
{\hfill$\diamondsuit$}
\end{remark}

\subsection{$L$-encounters}

Let $c$ be a $T$-periodic orbit involving an $L$-encounter ($L\geq 3$).
Without loss of generality, we assume that the encounter corresponds to the
trivial permutation $ e=\Big(\scriptsize \begin{array}{cccc}1&2&\dots&L\\1&2&\dots&L\end{array}\Big)$
and its orbit loops correspond to the permutation $P_{\rm loop}=\Big(\scriptsize\begin{array}{cccc}1&2&\dots&L\\2&3&\dots&1\end{array}\Big)$; recall section \ref{combisec}.
Let $P$ be a permutation in $S_L$ such that $P_{\rm loop} P$ is a single cycle. In this subsection, we will construct the partner orbit given by $P$. We define 
the {\em sequence $\{P_k\}_{k=0,\dots,L-3}$ generated by $P$} as follows.  Put $P_0:=P$ and for $k\in\{1,\dots,L-3\}$, define
$P_{k}:\{1,2,\dots,L\}\setminus\{2,\dots,k+1\}\rightarrow\{1,2,\dots,L\}\setminus\{1,\dots,k\}$  recursively by
\setlength{\parskip}{-0.2cm} 
\begin{equation}\label{Pkdf}   
P_{k}(j)=\begin{cases} P_{k-1}(j)&\quad\mbox{if}\quad j\ne P^{-1}_{k-1}(k)\\
 P_{k-1}(k+1) &\quad \mbox{if}\quad j=P^{-1}_{k-1}(k);
\end{cases}
\end{equation}
the respective orbit loops  
$P_{k,\rm loop}:\{1,2,\dots,L\}\setminus\{1,\dots,k\}\rightarrow\{1,2,\dots,L\}\setminus\{2,\dots,k+1\}$
is  defined by
\[ P_{k,\rm loop}(j)=
\begin{cases}
 1 &\quad\mbox{if}\quad j=L,\\
 j+1&\quad \mbox{otherwise}.
\end{cases} \] 
Then $P_{k,\rm loop} P_k$ is a permutation of the set $\{1,2,\dots,L\}\setminus\{2,\dots,k+1\}$ and is a single cycle.
Denote
\begin{eqnarray}\label{Deltad}
\Delta d_{(L)}&=& 41\Big(\frac{1}{3}+\cdots+\frac1L\Big),\quad \quad \ \,
\Delta T_{(L)}\ \,= \ \, 14+78\Big( \frac{1}{3^2}+\cdots+\frac{1}{L^2}\Big),
\\ 
\alpha_L&=&  6+468\Big(\frac{1}{3^3}+\cdots+\frac{1}{L^3}\Big),\ \ \ 
\beta_L\ \, =\ \, 1+17\Big(\frac{1}{3}+\cdots+\frac{1}{L}\Big),
\end{eqnarray}
 and recall $\eps_*$ from Lemma \ref{per-coinc}. The main result of this paper is the following.
\begin{theorem}\label{thmL} For $\eps\in\,]0, \frac{\eps_*}{\Delta d_{(L)}}]$,
 assume that the $T$-\,periodic orbit $c$ has an $(L,\frac{\eps}L)$-encounter with the following property:
\begin{enumerate}
\item[(i)] there are $x_j\in c$, $x_j\in\P_{\frac{\eps}L}(x)$,
$x_j=(u_j,s_j)_{x}$ for $j=1,\dots,L$ and some $ x\in X$;
\item[(ii)] $|u_j-u_i|>\frac{6}5\,(e^{-T_j}+e^{-T_i})$
 and $|s_j-s_i|>\frac{24}{L^3}\,\eps^3+e^{-T_{j-1}}+e^{-T_{i-1}}$ for $ j\ne i$;
here $T_1,\dots,T_L>0,\, T=T_1+\cdots+T_L$,\, $\varphi_{T_j}(x_j)=x_{j+1}$
for $j=1,\dots,L-1$ and $\varphi_{T_L}(x_L)=x_1$.
\end{enumerate}
Then for every $P\in S_L$ such that $P_{\rm loop} P$ is a single cycle,
the orbit $c$ has a $\Delta d_{(L)}\eps$-partner $c'$ given by $P$ with period $T'$ satisfying
\begin{equation}\label{actiondiff}
 \Big|\frac{T'-T}{2}-\Delta S_L\Big|\leq  \omega_L\eps^4
+ \kappa_L\eps^2(e^{-T_1}+\cdots+e^{-T_L}),
\end{equation}
where 
\vskip-0.5cm\[\Delta S_L=\sum_{j=1}^{L-2}\ln(1+(u_{j+1}-u_j)(s_{j+1}-s_1))
+\sum_{j=1}^{L-2}\ln(1+(u_{P_{j-1}(j+1)}-u_j)(s_{P_{j-1}^{-1}(j)}-s_{j+1})),\]
\begin{eqnarray}\label{omegakappa,L}
\omega_L & = & 12\Big(\frac{1}{3^4}+\cdots+\frac{1}{L^4}\Big)
+720\Big(\frac{1}{3^3}+\cdots +\frac{1}{L^3}\Big)
+\frac{21\alpha_L}{L}-114,
\\
\kappa_L & = & 118\Big(\frac{1}{3^2}+\cdots+\frac{1}{L^2}\Big)
+312\Big(\frac{1}{3}+\cdots+\frac{1}{L}\Big)+\frac{21 \beta_L}{L}-156. 
\end{eqnarray}
Furthermore, the partner orbit $c'$ has an $(L,\frac{3\eps}L)$-encounter. More precisely, there are
$v_1,\dots, v_L\in  c', T_1',\dots,T_L'>0$ such that for 
$j=1,\dots,L$:
\begin{enumerate}
\item[(a)] $\varphi_{T'_{P(j)}}(v_j)=
\begin{cases} v_{P(j)+1}&\mbox{if}\quad P(j)\ne L\\
v_1&\mbox{if}\quad P(j)=L
\end{cases}$\quad and\quad $T'=T_1'+\cdots+T_L'$;
\item[(b)]  $v_j\in \P_{\frac{3\eps}L} (x)$ for $v_j=(u'_j,s'_j)_x$ satisfying
\begin{subequations}\label{u's'L}
\begin{alignat}{2}
\label{uj'L}
|u_j'-u_{P(j)}|
&\,<\,\alpha_L\eps^3+\beta_L\eps(e^{-T_1}+\cdots+e^{-T_L}),
\\ \label{sj'L}
|s_j'-s_j|
&\,<\, \alpha_L\eps^3+\beta_L\eps(e^{-T_1}+\cdots+e^{-T_L});
\end{alignat}
\end{subequations}
\item[(c)] $|T'_j-T_j|<\Delta T_{(L)}\eps^2$;
\item[(d)] $d_X(\varphi_{t}(x_{P(j)}),\varphi_{t}(v_j))< \Delta d_{(L)}\eps$\quad for all\quad
$t\in \big[0,\max\{T_{P(j)},T'_{P(j)}\}\big]$.
\end{enumerate}

\end{theorem}
\setlength{\parskip}{0cm} 

\noindent{\bf Proof.}
Let $L\geq 3$ be an integer number. We see that
$\Delta T_{(L)}=\Delta T_{(L-1)}+\frac{78}{L^2}$ and $\Delta d_{(L)}=\Delta d_{(L-1)}+\frac{41}{L}$.
  We will prove this theorem by induction.
For $L=3,$  only one permutation
$P=\Big(\scriptsize\begin{array}{ccc}1&2&3\\ 2&3&1\end{array}\Big)$ satisfies the assumption, and 
 Theorem \ref{3-encounter} proves this case; note that
 $\omega_3>76,\kappa_3>7,\alpha_3> 23$, and $\beta_3>6$. 
Assume that the theorem is correct for $L=n-1$ for $n\geq 5$, we prove that it is correct for $L=n$.

Let $c$ be a periodic orbit with $n$-encounter illustrated by
the trivial permutation and let $P\in S_n$ be a permutation such that  $P_{\rm loop} P$ is a single cycle.
 Suppose that there are
$x\in X,\, x_1,\dots, x_n\in c$, $x_j=(u_j,s_j)_x$  satisfying the assumption. 
 Using Lemma \ref{lm1}, write 
$\tilde x_j:=\varphi_{\tilde \tau_j}(x_j)\in\P_{\frac{3\eps}{n}}(x_1)$ with $\tilde x_j=(\tilde u_j,\tilde s_j)_{x_1}$, and
\setlength{\parskip}{-0.1cm} 
\begin{align}\label{tildeL-}
|\tilde u_j-(u_j-u_1)|<\frac{8}{n^3}\,\eps^3,\quad
|\tilde s_j-(s_j-s_1)|<\frac{8}{n^3}\,\eps^3, \quad
|\tilde\tau_j|<\frac{8}{n^2}\,\eps^2, 
\end{align}\vskip-0.8cm
\begin{align}
\label{tildeujtildesj}
|\tilde u_j\tilde s_j-(u_j-u_1)(s_j-s_1)|<\frac{4}{n^4}\,\eps^4,
\quad  
 |\tilde u_j\tilde s_j|<\frac{5}{n^2}\,\eps^2.
\end{align}
Denoting $\tilde T_1= T_1+\tilde \tau_2$, $\tilde T_j=T_j-\tilde \tau_{j}+\tilde\tau_{j+1}$,
 for $j=2,\dots,n-1$, and $\tilde T_{n}=T_{n}-\tilde \tau_{n}$, we obtain
  $T=\tilde T_1+\cdots+\tilde T_{n}$,
\begin{equation}\label{tildexL}
\varphi_{\tilde T_j}(\tilde x_j)=\tilde x_{j+1}\quad \mbox{for} \quad j=1,\dots,n-1 
\quad \mbox{and}
\quad
\varphi_{\tilde T_{n}}(\tilde x_{n})=\tilde x_1=x_1.
\end{equation} 
Furthermore,
\begin{eqnarray}
\label{tildeT1-T1}
|\tilde T_1-T_1|=|\tilde\tau_2|< \frac{8}{n^2}\,\eps^2,\quad 
|\tilde T_j-T_j|\leq\frac{12}{n^2}\,\eps^2
\ \mbox{for}\ j=2,\dots,n-1,\quad 
|\tilde T_n-T_n|=|\tilde \tau_n|<\frac{8}{n^2}\,\eps^2.
\end{eqnarray}

Figure \ref{induct} below illustrates the idea of the proof. The encounter of the original orbit corresponds the trivial permutation depicted in (a).  In (b), we exchange the ports of the first two stretches to have two shorter periodic orbits $c_1$ and $c_2$ (Step 1 below). The longer one $c_1$ has $(n-1)$-encounter. We construct the partner orbit $c_1'$ corresponding the permutation $P_1$ for the $c_1$, expressed in (c) (Step 2). Finally, we connect two orbits to have a new partner orbit described in (d) (Step 3).

\begin{figure}[ht]
	\begin{center}
		\begin{minipage}[ht]{1\linewidth}
			\centering
			\includegraphics[angle=0,width=1\linewidth]{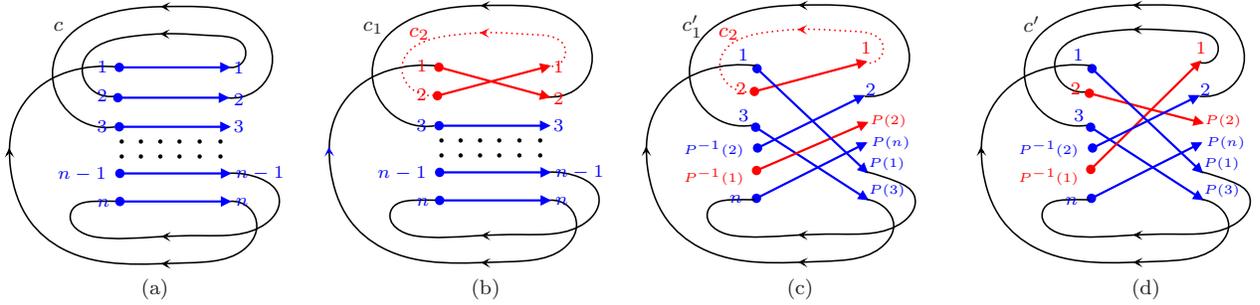}
		\end{minipage}
	\end{center}
	\caption{Inductive argument to construct partner orbits}\label{induct}
\end{figure}

\underline{Step 1:} {\em Reducing the encounter order}.  
Write $x_j=\Gamma g_j $ and $\tilde x_j=\Gamma \tilde g_j$ for
$g_j,\tilde g_j\in\G, j=1,\dots,n$.
By the above setting, 
$\tilde x_2=\varphi_{\tilde T_1}(x_1)\in\P_{\frac{3\eps}{n}}(x_1)$ with 
$\varphi_{\tilde T_1}(x_1)=( \tilde u_2,\tilde s_2)_{x_1}$. Note that  assumption (ii)  and \eqref{tildeT1-T1} guarantee that $\tilde T_1\ge 1$.  
 Then by the Anosov closing lemma I, there are
$\hat T_1\in\R$ and $y_2=\Gamma g_1c_{\sigma_2} b_{\eta_2}$ so that $\varphi_{\hat T_1}(y_2)=y_2$,
\begin{eqnarray}
\label{eta2-}
|\eta_2-\tilde s_2|
<\frac{30}{n^3}\,\eps^3+\frac{6}{n}\,\eps e^{-\tilde T_1},\quad 
|\sigma_2|<
2|\tilde u_2|e^{-\tilde T_1}<\frac{6}{n}\,\eps e^{-\tilde T_1},
\end{eqnarray}
\begin{equation*}
d_X(\varphi_t(x_1),\varphi_t(y_2))<2|\tilde u_2|+|\eta_2|\quad\mbox{for all}\quad t\in[0,\tilde T_1],
\end{equation*}
and 
\begin{equation*}
 \Big|\frac{\hat T_1-\tilde T_1}{2}-\ln(1+ \tilde u_2\tilde s_2)\Big|\leq 5| \tilde u_2\tilde s_2|e^{-\tilde T_1}.
\end{equation*}
Then
$
|\hat T_1-\tilde T_1|<\frac{21}{n^2}\,\eps^2$
implies that 
\begin{equation}\label{hatT1-T1}
|\hat T_1-T_1|<|\hat T_1-\tilde T_1|+|\tilde T_1-T_1|<\frac{29}{n^2}\,\eps^2
\end{equation}
due to \eqref{tildeT1-T1}. 
Furthermore,  using \eqref{tildeujtildesj}, we have
\begin{eqnarray*}
|\ln(1+\tilde u_2\tilde s_2)-\ln(1+(u_2-u_1)(s_2-s_1))|
<3|\tilde u_2\tilde s_2-(u_2-u_1)(s_2-s_1)|
<\frac{12}{n^4}\,\eps^4,
\end{eqnarray*}
owing to $|\ln(1+x)-\ln(1+y)|<3|x-y|$ for $x,y\in [0,\frac12]$;  
and so,
\begin{eqnarray}\label{T1}
\Big|\frac{\hat T_1-\tilde T_1}{2}-\ln(1+(u_2-u_1)(s_2-s_1))\Big|
< \frac{12}{n^4}\,\eps^4+\frac{26}{n^2}\,\eps^2e^{-T_1}.
\end{eqnarray}
It follows from \eqref{eta2-} and \eqref{tildeL-}
that 
\begin{eqnarray}\label{eta2+s1}
|\eta_2+s_1|
\leq |\eta_2-\tilde s_2| +|\tilde s_2-(s_2-s_1)|+|s_2|
<\frac{2}{n}\,\eps,
\end{eqnarray}
whence $
|\eta_2|<\frac{3}{n}\,\eps$,
and consequently
\begin{equation}\label{x1y2}
d_X(\varphi_t(x_1),\varphi_t(y_2))<\frac{9}{n}\,\eps \quad\mbox{for all}\quad t\in[0,\tilde T_1].
\end{equation}
The orbit of $y_2$  is depicted by the dotted line $c_2$ in Figure \ref{induct}\,(b).
\medskip

Similarly, $\varphi_{T-\tilde T_1}(\tilde x_2)=x_1
=\Gamma  \tilde g_2 b_{-\tilde s_2}c_{- \tilde u_2}\in\P'_{\frac{3\eps}{n}}(\tilde x_2)$
 with $\varphi_{T-\tilde T_1}(\tilde x_2)=(-\tilde s_2,-\tilde u_2)_{\tilde x_2}'$.
  Applying the Anosov closing lemma II, there are $y_1=\Gamma \tilde g_2 b_{\eta_1}c_{\sigma_1}$
and $\check T\in\R$ so that $\varphi_{\check T}(y_1)=y_1,$
\begin{equation}\label{checkT}
 \Big|\frac{\check T-({T-\tilde T_1})}{2} \Big|<4| \tilde u_2\tilde s_2|e^{-T+\tilde T_1}
 <\frac{1}{n^2}\,\eps^2 e^{-T_2},
\end{equation}
\begin{equation*}
d_X(\varphi_t(y_1),\varphi_t(\tilde x_2))
<2|\tilde u_2|+|\eta_1|\quad \mbox{for all}\quad t\in [0,T-\tilde T_1],
\end{equation*}
and
\begin{equation*}\label{sigma1eta10}
|\sigma_1|<\frac{6}{n}\,\eps e^{-(\tilde T_2+\cdots+\tilde T_{n})},\quad
|\tilde s_2+\eta_1|<\frac{6}{n}\,\eps e^{-(\tilde T_2+\cdots+\tilde T_{n})}.
\end{equation*}
Then 
\begin{equation}\label{sigma1eta1}
|\sigma_1|<\frac{7}{n}\,\eps e^{-(T_2+\cdots +T_{n})},\quad
|\tilde s_2+\eta_1|<\frac{7}{n}\,\eps e^{-(T_2+\cdots +T_{n})}
\end{equation}
and \eqref{tildeL-} yield 
\begin{eqnarray*}
|\eta_1|
\leq |\tilde s_2+\eta_1|+ |\tilde s_2-(s_2-s_1)|+|s_2-s_1|
<\frac{3}{n}\,\eps;
\end{eqnarray*}
so that
\begin{equation}\label{y1tildex2}
d_X(\varphi_t(y_1),\varphi_t(\tilde x_2))<\frac{9}{n}\,\eps\quad \mbox{for all}\quad t\in [0,T-\tilde T_1].
\end{equation}
The orbit of $y_1$ is expressed by the solid line $c_1$ in Figure \ref{induct}\,(b).

\medskip 
In order to apply the inductive assumption, it is necessary to verify that the orbit $c_1$ through
 $y_1$ has an $(n-1)$-encounter.
Indeed, recall that 
$y_1=\Gamma  \tilde g_2b_{\eta_1}c_{\sigma_1}
=\Gamma g_1c_{\tilde u_2}b_{\tilde s_2}b_{\eta_1}c_{\sigma_1}=
\Gamma g c_{u_1}b_{s_1} c_{\tilde u_2}b_{\tilde s_2+\eta_1}c_{\sigma_1}$.
 Apply Lemma \ref{bcbcbc} to write $y_1=\Gamma g c_{\check u_1} b_{\check s_1} a_{\check\tau_1}$
for
\begin{eqnarray*}
\check u_{1}&=&u_1+\tilde u_2+\sigma_1
+\frac{1}{1+\check\rho_1}\,\big({\tilde u_2\sigma_1(\tilde s_2+\eta_1)-(\tilde u_2+\sigma_1)\check\rho_1}\Big),
\\
\check s_{1}&=&s_1+\tilde s_2+\eta_1+\check\rho_1(s_1+\tilde s_2+\eta_1)+\tilde u_2s_1(\tilde s_2+\eta_1)(1+\check\rho_1)
, \\
 \check\tau_{1}&=& 2\ln(1+\check\rho_1),
\end{eqnarray*}
where \[\check\rho_1=\sigma_1(s_1+\tilde s_2+\eta_1)+\tilde u_2s_1(1+\sigma_1(\tilde s_2+\eta_1)).\] 
Using \eqref{sigma1eta1}, we have 
$
|\check\rho_1|
<\frac{4}{n^2}\,\eps^2$.
This implies that
\begin{eqnarray}\notag 
 |\check u_1-u_2|
 &\leq& |\check u_1- (u_1+\tilde u_2)| +|\tilde u_2-(u_2-u_1)|
 \\  \label{checku1-}
  &<&\frac{28}{n^3}\,\eps^3+\frac{8}{n}\,\eps e^{-(T_2+\cdots+T_n)},
 \\ \label{checks1-}
|\check s_1-s_1|&<& 
 \frac{2}{n^3}\,\eps^3,
\\ \label{checktau1}
|\check\tau_1|&<& \frac{16}{n^2}\,\eps^2.
\end{eqnarray}
For $j=3,\dots,n$, define $y_j:=\varphi_{-\tilde\tau_2+T_2+\cdots T_{j-1}}(y_1)$. Then, using
$\varphi_{T_2+\cdots+T_{j-1}}(x_2)=x_j=(u_j,s_j)_x$ and Lemma \ref{bcbcbc}, we write
\begin{eqnarray*} 
y_j & = & \Gamma \tilde g_2 
a_{-\tilde\tau_2+T_2+\cdots+T_{j-1}}
b_{\eta_1 e^{-(-\tilde\tau_2+T_2+\cdots+T_{j-1})}}
c_{\sigma_1 e^{-\tilde\tau_2+T_2+\cdots+T_{j-1}}}
\\
 & = &\Gamma g_j b_{\eta_1 e^{-(-\tilde\tau_2+T_2+\cdots+T_{j-1})}}
 c_{\sigma_1 e^{-\tilde\tau_2+T_2+\cdots+T_{j-1}}}
 \\
 & = &\Gamma  g c_{u_j}b_{s_j+\eta_1 e^{-(-\tilde\tau_2+T_2+\cdots+T_{j-1})}}
 c_{\sigma_1 e^{-\tilde\tau_2+T_2+\cdots+T_{j-1}}}
 \\
 & = &\Gamma g c_{\check u_j}b_{\check s_j}a_{\check\tau_j}
\end{eqnarray*} 
for 
\begin{eqnarray*}
		\check u_j & = & u_j+ \frac{{\sigma_1e^{-\tilde\tau_2+T_2+\cdots+T_{j-1}}}}{1+\check\rho_j },
	\\
	\check s_j & = & s_j+\eta_1 e^{-(-\tilde\tau_2+T_2+\cdots+T_{j-1})}
	+ \check\rho_j(s_j+\eta_1 e^{-(-\tilde\tau_2+T_2+\cdots+T_{j-1})}),\\
	\check\tau_j & = & 2\ln(1+\check\rho_j);
\end{eqnarray*}
here $\check\rho_j= s_j\sigma_1 e^{-\tilde\tau_2+T_2+\cdots+T_{j-1}}+\sigma_1\eta_1$. 
Recall that $|\sigma_1|<\frac{6}{n}\eps e^{\tilde \tau_2-(T_2+\cdots+T_n)},
\, |\eta_1|<\frac{3}{n}\eps$ and $|\tilde \tau_2|<\eps^2$.
Then $|\check\rho_j|< \frac{10}{n^2}\eps^2 e^{-(T_j+\cdots+T_n)}$
implies 
\begin{eqnarray}\label{checktauj}
|\check\tau_j|&<&\frac{1}{n^2}\,\eps^2,
\\ 
|\check u_j-u_j| &<& \frac{8}n\,\eps e^{-(T_j+\cdots + T_n)},
\\ 
|\check s_j-s_j| &<&  \frac{4}{n}\,\eps e^{-(T_j+\cdots+T_n)}. 
\end{eqnarray}
Together with \eqref{checku1-} and \eqref{checks1-} we obtain
\begin{subequations}\label{checkujsj}
	\begin{alignat}{2}\label{checkuj-}
	|\check u_1-u_2|, |\check u_j-u_j| &<\frac{28}{n^3}\,\eps^3+ \frac{8}n\,\eps e^{-(T_j+\cdots + T_n)}, \quad j=3,\dots,n,
	\\ \label{checksj-}
	|\check s_j-s_j| &< \frac{2}{n^3}\,\eps^3+ \frac{4}{n}\,\eps e^{-(T_j+\cdots+T_n)},\quad j=1,3,\dots,n, 
	\end{alignat} 
\end{subequations}
and as a consequence $|\check s_j|<\frac{1}{n-1}\,\eps $ as well as $|\check u_j|<\frac{1}{n-1}\,\eps$, for $j=1,3,\dots,n$. 
Therefore,  
$\check y_{j}:=\varphi_{-\check\tau_j}(y_j)
\in\P_{\frac{\eps}{n-1}}(x)$ with $\check y_j=(\check u_j,\check s_j)_{x}$ 
for $j=1,3,\dots,n$.
In addition, letting
\begin{eqnarray*}
\check T_2&=&-\tilde{\tau}_2+T_2+\check \tau_{1}-\check\tau_3,
\\
 \check T_j&=& T_j+\check\tau_{j}-\check\tau_{j+1}\quad \mbox{for}\quad j=3,\dots,n-1,
 \\
\check T_{n}&=&\check T-(\check T_2+\check T_3+\cdots+\check T_{n-1}),
 \end{eqnarray*}
we have $\check T= \check T_2+\cdots+\check T_n$ and
$\varphi_{\check T_2}(\check y_1)=\check y_3, \varphi_{\check T_{j}}(\check y_j)
=\check y_{j+1}$ for $j=3,\dots,n-1$ and $\varphi_{\check T_{n}}(\check y_{n})=\check y_1$. 
This means that the orbit $c_1$ has an $(\frac{\eps}{n-1},n-1)$-encounter.

\medskip 
Next we derive an  estimate for $|\check T_j-T_j|$ that will be helpful. A short calculation shows that
\begin{subequations}\label{Tj-hatTj}
\begin{alignat}{2}
T_2-\check T_2
&\ =\ \tilde \tau_2-\check \tau_1 +\check \tau_3,
\\
T_j-\check T_j
&\ =\ \check \tau_{j+1}-\check \tau_j,\quad \mbox{for}\quad j=3,\dots,n-1,
\\
T_n-\check T_n
&\ =\ T-\tilde T_1-\check T_1+\check \tau_1-\check\tau_n.
\end{alignat}
\end{subequations}
By \eqref{tildeL-}, \eqref{checktau1},\eqref{checktauj},
and \eqref{checkT}, we have
\begin{eqnarray}
|\check T_2-T_2|< \frac{25}{n^2}\,\eps^2,\ \ 
|\check T_j-T_j|<\frac{2}{n^2}\,\eps^2,\ j=3,\dots,n-1,\ \ \mbox{and}\ \ 
|\check T_n-T_n|< \frac{19}{n^2}\,\eps^2. 
\end{eqnarray} 
In summary,
\begin{equation}\label{checkTj-}
|\check T_j-T_j|<\frac{25}{n^2}\,\eps^2\quad\mbox{for}\quad j=2,\dots,n.
\end{equation}

\smallskip 
\underline{Step 2:} {\em Applying the inductive assumption}.
The orbit $c_1$ is expressed by the bijection 
$\hat P: \{1,3,\dots,n\}$ $\rightarrow$ $\{2,3,\dots,n\}$ defined by
$\hat P=\scriptsize\Big( \begin{array}{ccccccc} 1&3&\cdots & n
\\ 2 &3 &\cdots &n\end{array}\Big)$; see the solid line in Figure \ref{induct}\,(b).
The orbit loops of $c_1$ are illustrated by $\hat P_{\rm loop}=P_{1,\rm loop}=\scriptsize\Big( \begin{array}{ccccccc} 2&3&\cdots & n-1 & n
\\ 3 &4 &\cdots &n& 1\end{array}\Big) $.  
Let $\check P=P_1$ be defined by \eqref{Pkdf}:
\[\check P(j)
=\begin{cases} P(j) \quad \mbox{if}\quad j\ne P^{-1}(1),
\\
P(2) \quad\mbox{if}\quad j=P^{-1}(1).
\end{cases}\]
 Then $\hat P_{\rm loop}\check P$ is a permutation of the set $\{2,3,\dots,n\}$ and is a single cycle
 since $P_{\rm loop}P$ is a single cycle of the set
 $\{1,2,\dots,n\}$. 
By the inductive assumption, the orbit $c_1$ has a partner orbit $c_1'$ which is illustrated by $\check P$;
see the solid line in Figure \ref{induct}\,(c). 
We see that the sequence $\{\check P_k\}_{k=0,\dots,n-4}$ generated by 
$\check P$ satisfies
$\check P_{k}=P_{k+1}$ for $k=0,\dots,n-4$;
recall the definition of $P_k$ in \eqref{Pkdf}.  
Let $\hat T$ be the period of $c_1'$, then 
\begin{eqnarray}\label{hatTap}
\Big|\frac{\hat T-\check T}{2}-\Delta \check S_{n-1}\Big|
<\omega_{n-1}\eps^4+ \kappa_{n-1}\eps^2(e^{-\check T_2}+\cdots+e^{-\check T_{n}}),
\end{eqnarray} 
where
\begin{eqnarray*}
\Delta \check S_{n-1}
 &=&\ln(1+(\check u_3-\check u_1)(\check s_3-\check s_1))
 +\sum_{j=3}^{n-2}\ln(1+(\check u_{j+1}-\check u_j)(\check s_{j+1}-\check s_1))
 \\
 &&+\,\ln(1+(\check u_{\check P(3)}-\check u_1)(\check s_{\check P^{-1}(2)}-\check s_3))
 +\, \sum_{j=2}^{n-3}\ln(1+(\check u_{\check P_{j-1}(j+2)}-\check u_{j+1})(\check s_{\check P_{j-1}^{-1}(j+1)}-\check s_{j+2}))
 \\
 &=&\ln(1+(\check u_3-\check u_1)(\check s_3-\check s_1))
 +\sum_{j=3}^{n-2}\ln(1+(\check u_{j+1}-\check u_j)(\check s_{j+1}-\check s_1))
 \\
 &&+\,\ln(1+(\check u_{P_1(3)}-\check u_1)(\check s_{P_1^{-1}(2)}-\check s_3))
+\sum_{j=3}^{n-2}\ln(1+(\check u_{P_{j-1}(j+1)}-\check u_j)(\check s_{P_{j-1}^{-1}(j)}-\check s_{j+1}))
\end{eqnarray*}
and $\omega_{n-1}, \kappa_{n-1}$ are defined by \eqref{omegakappa,L}. 
Denote 
\[\Delta S_{n-1}=\sum_{j=2}^{n-2}\ln(1+( u_{j+1}- u_j)(s_{j+1}-s_1))+
\sum_{j=2}^{n-2}\ln(1+( u_{P_{j-1}(j+1)}- u_j)(s_{P_{j-1}^{-1}(j)}-s_{j+1})).\]
For $j,i,m,l\in \{1,3,\dots,n\}$, using \eqref{checkujsj}, 
we can show that 
\begin{eqnarray}
|(\check u_j-\check u_i)(\check s_l-\check s_m)-(u_j-u_i)(s_l-s_m)|
<\frac{120}{(n-1)n^3}\,\eps^4+\frac{52}{n(n-1)}\,\eps^2 e^{-T_n}.
\end{eqnarray}
Then 
\begin{align*}
&\hskip-1cm |\Delta \check S_{n-1}-\Delta S_{n-1}|
\leq 
3\,|(\check u_3-\check u_1)(\check s_3-\check s_1)
-(u_3-u_2)(s_3-s_1)|
\\
&\,+3\,|(u_{P_1(3)}-u_2)(s_{P_1^{-1}(2)}-s_3) -(\check u_{P_1(3)}-\check u_1)(\check s_{P_1^{-1}(2)}-\check s_3)|
\\
&\,+\,
3\sum_{j=3}^{n-2}|( u_{j+1}- u_j)(s_{j+1}-s_1)-(\check u_{j+1}-\check u_j)(\check s_{j+1}-\check s_1)|
\\
&\,+\,3\sum_{j=3}^{n-2}|(\check u_{P_{j-1}(j+1)}-\check u_j)
( \check s_{P_{j-1}^{-1}(j)}-\check s_{j+1})-(u_{P_{j-1}(j+1)}- u_j)(s_{P_{j-1}^{-1}(j)}-s_{j+1})|
\\
&\,\leq\, 2(n-3)\cdot 3\Big(\frac{120}{(n-1)n^3}\,\eps^4+\frac{52}{n(n-1)}\,\eps^2 e^{-T_n}\Big)
\,\leq \, \frac{720}{n^3}\,\eps^4+\frac{312}{n}\,\eps^2 e^{-T_n}
\end{align*}
implies 
\begin{eqnarray} \label{T3} \notag 
\Big|\frac{\hat T-\check T}2-\Delta S_{n-1}\Big|
&\leq&\Big|\frac{\hat T-\check T}{2}-\Delta \check S_{n-1}\Big|
+|\Delta \check S_{n-1}-\Delta S_{n-1}|
\\
&\leq& \Big(\omega_{n-1}+\frac{720}{n^3}\Big) \eps^4+
\Big(\kappa_{n-1}+\frac{312}{n}\Big) \eps^2( e^{-T_1}+\cdots+e^{-T_{n}})
\end{eqnarray}
due to \eqref{hatTap}. 
\smallskip

Also by the inductive assumption, the partner orbit $c_1'$ has an $(n-1,\frac{3\eps}{n-1})$-encounter.
More precisely, there are $z_1,z_3,\dots,z_{n}\in c_1',z_j\in\P_{\frac{3\eps}{n-1}}(x)$ 
for $z_j=(\hat u_j,\hat s_j)_x$ and  $\hat T_2,\dots,\hat T_{n}>0$ such that
\begin{eqnarray}
\label{hatTj-checkTj}
|\hat T_j-\check T_j|&\leq&\Delta T_{(n-1)}\eps^2,
\\
\label{inductive}
\varphi_{\hat T_{P_1(j)}}(z_j)&=&
\begin{cases}
z_{P_1(j)+1}&\mbox{if}\quad P_1(j)\ne n\\
z_1&\mbox{if}\quad P_1(j)=n,
\end{cases}
\end{eqnarray}
 \begin{equation}
 \label{d(n-1)}
 d_X(\varphi_{t}(\check y_{P_1(j)}),\varphi_{t}(z_j))<\Delta d_{(n-1)}\eps
  \quad\mbox{for all}\quad  t\in [\,0,\,\max\{\check T_{P_1(j)},\hat T_{P_1(j)}\}\,],
 \end{equation}
and 
\begin{subequations} \label{checkusj,L}
\begin{alignat}{2}
|\hat u_j-\check u_{P_1(j)}|
\ <\ & \alpha_{n-1}\eps^3+ \beta_{n-1}\eps(e^{- \check T_2}+\cdots+e^{-\check T_n}),
\\
|\hat s_j-\check s_j|\ <\ & \alpha_{n-1}\eps^3+\beta_{n-1}\eps (e^{-\check T_2}+\cdots+e^{-\check T_{n}}).
\end{alignat}
\end{subequations} 
with the convention
\begin{equation}\label{convention}
\check y_2= \check y_1 \quad \mbox{as well as}\quad 
\check u_2=\check u_1.
\end{equation}
Recall that $\tilde \tau_2=-2\ln(1-s_1(u_2-u_1)),
\check \tau_j=2\ln(1+\check\rho_j)$.
Furthermore, according to the proof of the Anosov closing lemma II,
$T-\tilde T_1-\check T_1=2\ln(1+(\tilde s+\eta_1)\tilde u_2)$. 
Whence, it follows from \eqref{Tj-hatTj} that 
\begin{eqnarray*}
e^{T_2-\check T_2}
&=&(1-s_1(u_2-u_1))^{-2}(1+\check \rho_1)^{-2}(1+\check\rho_3)^2,
\\
e^{T_j-\check T_j}&=&(1+\check\rho_{j+1})^2(1+\check\rho_j)^{-2}, \ 
j=3,\dots,n-1,
\\
e^{T_n-\check T_n}
&=&(1+(\tilde s_2+\eta_1)\tilde u_2)^2(1+\check \rho_1)^2(1+\check\rho_n)^{-2}.
\end{eqnarray*}
Then 
\begin{equation*}
e^{T_2-\check T_2} \leq (1+4|s_1(u_2-u_1)|)(1+4|\check \rho_3-\check \rho_1|) <1+ \frac{22}{n^2}\,\eps^2.
\end{equation*}
Similarly, we obtain
\[e^{T_j-\check T_j}<1+\frac{1}{n^2}\,\eps^2  
\quad\mbox{for}\quad j=3,\dots,n-1\quad \mbox{and}\quad e^{T_n-\check T_n} <1+\frac{18}{n^2}\,\eps^2.\]
Therefore, it follows from \eqref{checkusj,L} that
\begin{eqnarray*}
|\hat u_j-\check u_{P_1(j)}|
&<& \alpha_{n-1}\eps^3+ \Big(\beta_{n-1}+\frac{1}{n}\Big)\eps(e^{-T_2}+\cdots+e^{-T_n})
\\
|\hat s_j-\check s_j| &< & \alpha_{n-1}\eps^3+\Big(\beta_{n-1}+\frac{1}{n}\Big)\eps (e^{-T_2}+\cdots+e^{- T_{n}});
\end{eqnarray*}
note that $\beta_{n-1}\frac{22}{n^2}\eps^2<\frac{1}{n}$. 
Together with \eqref{checkujsj}, this shows that
\begin{subequations}\label{hatusj,L}
\begin{alignat}{2}\label{hatuj-}
|\hat u_j-u_{P_1(j)}|
&\ \leq  \Big(\alpha_{n-1}+\frac{28}{n^3}\Big)\eps^3+\Big(\beta_{n-1}+\frac{9}{n}\Big)\eps (e^{-T_2}+\cdots+ e^{-T_n}),
\\ \label{hatsj-}
|\hat s_j-s_j|&\ \leq\ \Big( \alpha_{n-1}+\frac{2}{n^3}\Big)\eps^3+\Big(\beta_{n-1}+\frac{5}{n} \Big)\eps (e^{-T_2}+\cdots+ e^{-T_n}).
  \end{alignat}
\end{subequations}

\smallskip 

\underline{Step 3:} {\em Construction of the partner $c'$ and proof of \eqref{actiondiff}}.
We use the connecting lemma to `connect' the orbit $c_1'$ and the orbit of $y_2=\Gamma h_2$
to obtain the new partner orbit $ c'$. To do this, we need to verify that the stretch $P^{-1}(1)$-th contains a point (called $\bar z_{P^{-1}(1)}$)
that lies on the Poincar\'e section of $y_2$.
Recall that $\Gamma g_1=\Gamma g c_{u_1}b_{s_1}$ and $\Gamma h_2=\Gamma g_1 c_{\sigma_2}b_{\eta_2}$.
 Applying Lemma \ref{bcbcbc}, we can write
 \begin{eqnarray*}
z_{P^{-1}(1)}
&=&\Gamma gc_{\hat u_{{P^{-1}(1)}}}b_{\hat s_{{P^{-1}(1)}}}
=\Gamma g_1c_{\sigma_2}b_{\eta_2}(b_{-\eta_2}c_{-\sigma_2}b_{-s_1}c_{-u_1+\hat u_{{P^{-1}(1)}}}b_{\hat s_{{P^{-1}(1)}}})
\\
&=&\Gamma h_2 c_{\bar u_{P^{-1}(1)}}b_{\bar s_{P^{-1}(1)}}a_{\bar \tau_{P^{-1}(1)}}
\end{eqnarray*}
and a short estimate shows that
\begin{eqnarray}
\label{bartauP-11}
|\bar \tau_{P^{-1}(1)}|
&<& 4|\bar\rho_{P^{-1}(1)}|<\frac{44}{n^2}\,\eps^2,
\\ \label{baru-0}
|\bar u_{P^{-1}(1)}-(-u_1+\hat u_{P^{-1}(1)})|
	 &<& \frac{72}{n^3}\,\eps^3+\frac{8}n\, \eps e^{-T_1},
\\ 
\label{bars-0}
|\bar s_{P^{-1}(1)}-(-\eta_2-s_1+\hat s_{P^{-1}(1)})|
&<&
\frac{100}{n^3}\,\eps^3.
\end{eqnarray}
So that  $|\bar s_{P^{-1}(1)}|<\frac{6\eps}n,|\bar u_{P^{-1}(1)}|<\frac{5\eps}n$ 
as well as \[\bar z_{P^{-1}(1)}:=\varphi_{-\bar\tau_{P^{-1}(1)}}(z_{P^{-1}(1)})=\Gamma h_2 c_{u_{P^{-1}(1)}}b_{s_{P^{-1}(1)}}\in \P_{\frac{6\eps}n}(y_2).\]
Now we can apply the connecting lemma (Lemma \ref{connect0}) to connect the $P^{-1}(1)$-th entrance port of $c_1'$
and the 1st exit port of the orbit $c_2$ through $y_2$; see the red thick lines in Figure \ref{induct}\,(c)\,\&\,(d). There are $T'\in\R,$
$w_{P^{-1}(1)}=\Gamma h_2c_{\bar u_{P^{-1}(1)}e^{-\hat T_1}+\sigma}b_\eta\in\P_{\frac{12\eps}n}(y_2)$
so that $\varphi_{T'}(w_{P^{-1}(1)})=w_{P^{-1}(1)}$,
\begin{eqnarray} \label{T'-(hatT1}
\notag
 \Big|\frac{T'-(\hat T_1+\hat T)}{2}-\ln(1+\bar u_{P^{-1}(1)}\bar s_{P^{-1}(1)})\Big|
&<& \frac{90}{n^2}\,\eps^2(e^{-\hat T_1}+ e^{-\hat T})+\frac{240}{n^2}\,\eps^2 e^{-\hat T_1-\hat T}
\\ 
&<& \frac{92}{n^2}\,\eps^2(e^{-T_1}+e^{-T_2}),
\end{eqnarray}
\vskip-0.7cm
\begin{eqnarray}
\label{phitwp-11}
  d_X(\varphi_t(w_{P^{-1}(1)}),\varphi_t(y_2))
  &<& \frac{30}{n}\, \eps
\quad\quad \mbox{for all}\quad t\in [0,\hat T_1],
  \\ \label{phitwp-112}
  d_X(\varphi_{t+\hat T_1}(w_{P^{-1}(1)}),\varphi_t(\bar z_{P^{-1}(1)}))
  &<&\frac{30}{n}\, \eps \quad\quad\mbox{for all}\quad t\in [0,\hat T],
\end{eqnarray} 
and \vskip -0.5cm
\begin{eqnarray}\label{eta-bars}
|\eta-\bar s_{P^{-1}(1)}|
<\frac{361}{n^3} \,\eps^3 ,\quad 
 |\sigma|< \frac{10}{n} \,\eps e^{-\hat T_1-\hat T}.
\end{eqnarray}
The partner orbit given by the permutation $P$ is illustrated in Figure \ref{induct}\,(d). 
Now we  prove \eqref{actiondiff}. First, we have 
\begin{eqnarray}\label{bars-}
\lefteqn{|\bar s_{P^{-1}(1)}-(s_{P^{-1}(1)}-s_2)|}
\notag \\ 
&\leq& |\bar s_{P^{-1}(1)}-(-\eta_2-s_1+\hat s_{P^{-1}(1)})|
+|\hat s_{P^{-1}(1)}-s_{P^{-1}(1)}|
+|\tilde s_2-\eta_2|+|(s_2-s_1)-\tilde s_2|
\notag\\ 
&\leq&\Big(\alpha_{n-1}+\frac{140}{n^3}\Big)\eps^3 +\Big(\beta_{n-1}+\frac{5}{n} \Big)\eps( e^{-T_1}+ \cdots+e^{-T_n}\Big)
\end{eqnarray}
due to \eqref{bars-0}, \eqref{hatsj-}, \eqref{eta2-}, and \eqref{tildeL-};
\begin{eqnarray}\label{baru-}
\hskip-1cm|\bar u_{P^{-1}(1)}-(u_{P(2)}-u_1)|
&\leq& |\bar u_{P^{-1}(1)}-(-u_1+\hat u_{P^{-1}(1)})|
+|\hat u_{P^{-1}(1)}-u_{P(2)}|
\notag \\
&\leq&\Big(\alpha_{n-1}+\frac{100}{n^3}\Big)\eps^3 +\Big(\beta_{n-1}+\frac{9}{n}\Big)\eps( e^{-T_1}+ \cdots+e^{-T_n}\Big)
\end{eqnarray}
using \eqref{hatuj-}, \eqref{baru-0} and noting that $P_1(P^{-1}(1))=P(2)$. This yields
\begin{eqnarray*}
&&\hskip -4cm \big|\ln(1+\bar u_{P^{-1}(1)}\bar s_{P^{-1}(1)})-\ln(1+(u_{P(2)}-u_1)(s_{P^{-1}(1)}-s_2))\big|
\\
&&< \frac{21\alpha_n}{n}\,\eps^4+\frac{21\beta_n}{n}\,\eps^2 (e^{-T_1}+\cdots+e^{-T_n});
\end{eqnarray*}
note that $\alpha_{n-1}+\frac{140}{n^3}<\alpha_n$ and
$\beta_{n-1}+\frac{9}{n}<\beta_n$. 
Together with \eqref{T'-(hatT1}, we obtain
\begin{eqnarray}\label{T4}
&& \hskip-4cm\Big|\frac{T'-(\hat T_1+\hat T)}{2}-\ln(1+(u_{P(2)}-u_1)(s_{P^{-1}(1)}-s_2))\Big|
\notag
\\ && \leq\  \frac{21\alpha_n}{n}\,\eps^4+\Big(\frac{21\beta_n}{n}+\frac{92}{n^2}\Big)\eps^2(e^{-T_1}+ \cdots+e^{-T_{n}}).
\end{eqnarray}
Now we are in a position to derive the action difference between $c$ and $c'$. Note that
\[\Delta S_{n}=\Delta S_{n-1}+\ln(1+(u_2-u_1)(s_2-s_1))+\ln(1+(u_{P(2)}-u_1)(s_{P^{-1}(1)}-s_2)).\]
By \eqref{T1}, \eqref{checkT}, \eqref{T3}, and \eqref{T4}, the action difference between $c$ and $c'$ satisfies
\begin{eqnarray*}
\lefteqn{\Big|\frac{T'-T}{2}-\Delta S_{n}\Big|
\leq \Big|\frac{\hat T_1-\tilde T_1}{2}-\ln(1+(u_2-u_1)(s_2-s_1))\Big|
+\Big| \frac{\check T-(T-\tilde T_1)}{2}\Big|}
\\
&&+\,\,\Big|\frac{\hat T-\check T}{2}-\Delta S_{n}\Big|
+\Big|\frac{T'-(\hat T_1+\hat T)}{2}-\ln(1+(u_{P(2)}-u_1)(s_{P^{-1}(1)}-s_2))\Big|\\
&&\leq\ \omega_n\eps^4+\kappa_n\eps^2(e^{-T_1}+\cdots+e^{-T_{n}});
\end{eqnarray*}
note that by \eqref{omegakappa,L}, we have
\begin{eqnarray*}
\omega_n=\omega_{n-1}+\frac{12}{n^4}+\frac{720}{n^3}+\frac{21\alpha_n}{n}\quad\mbox{and}\quad \kappa_n= \kappa_{n-1}+\frac{118}{n^2}+\frac{312}{n}+\frac{21\beta_{n}}{n}.
\end{eqnarray*}
 
\smallskip 

\underline{Step 4:} {\em Definition of $v_j,\,T_j'$
 and proof of (a)}. It has not been sufficient yet to prove that 
$|u_j'|<\frac{3}{n}\eps$ and $|s_j'|<\frac{3}{n}\eps$. We will do it after verifying (d). For simplicity, let us 
write $v_j\in \P_{\frac{3\eps}{n}}(x)$ if we derive
$v_j=\Gamma g c_{u'_j}b_{s'_j}$ for some $u'_j,s_j'\in\R$.  In what follows we often use the fact that 
$|\sigma_1|<\frac{7}n\eps e^{-(T_2+\cdots+T_n)}, |\eta_1|<\frac{3}{n}\eps,|\eta_2|<\frac{3}{n}\eps,
|\sigma_2|<\frac{6}{n}\eps e^{-\tilde T_1}, |\eta|<\frac{7}{n}\eps,
|s_1+\eta_2|<\frac{1}{n}\eps,
|\bar u_{P^{-1}(1)}|<\frac{5}{n}\eps, 
|\bar s_{P^{-1}(1)}|<\frac{6}{n}\eps
$. 

\medskip 
\noindent 
\underline{{Step 4.1:}} {\em  Definition of $v_{P^{-1}(1)}$.}
Recall  $w_{P^{-1}(1)}=\Gamma h_2 c_{\bar u_{P^{-1}(1)}e^{-\hat T_1}+\sigma} b_\eta$ 
and $\Gamma h_2=\Gamma g_1 c_{\sigma_2}b_{\eta_2}$. Then applying Lemma \ref{bcbcbc}, we can write
\begin{eqnarray*} 
w_{P^{-1}(1)}&=&
\Gamma g_1c_{\sigma_2}b_{\eta_2}c_{\bar u_{P^{-1}(1)}e^{-\hat T_1}+\sigma} b_\eta
=\Gamma g( c_{u_1}b_{s_1}c_{\sigma_2}b_{\eta_2}c_{\bar u_{P^{-1}(1)}e^{-\hat T_1}+\sigma} b_\eta)
\\
&=&\Gamma g c_{u'_{P^{-1}(1)}}b_{s'_{P^{-1}(1)}}a_{\tau'_{P^{-1}(1)}}
\end{eqnarray*}
and a short calculation shows
\begin{eqnarray}
\label{tau'p-11}
|\tau'_{P^{-1}(1)}|&<&\frac{1}{n^2}\,\eps^2,
\\ \label{u'P-11}
|u'_{P^{-1}(1)}-u_1|&<& 
 \frac{15}{n}\,\eps e^{-T_1},
 \\ \label{s'-s1}
 |s'_{P^{-1}(1)}-(s_1+\eta_2+\eta)|
 &<& \frac{5}{n^3}\, \eps^3 .
\end{eqnarray}
Together with \eqref{eta-bars}, \eqref{bars-0}, 
and \eqref{hatsj-},  we obtain
\begin{eqnarray}\notag 
|s'_{P^{-1}(1)}-s_{P^{-1}(1)}|
&\leq & |s'_{P^{-1}(1)}-(s_1+\eta_2+\eta)|
+|\eta-\bar s_{P^{-1}(1)}| 
\\ \notag 
&& +\ |\bar s_{P^{-1}(1)}-(-\eta_2-s_1+\hat s_{P^{-1}(1)})|
+  |\hat s_{P^{-1}(1)}- s_{P^{-1}(1)}|
\\ \notag 
&\leq&
\Big(\alpha_{n-1}+\frac{468}{n^3}\Big) \eps^3+\Big(\beta_{n-1}+\frac{5}{n} \Big)\eps (e^{-T_2}+\cdots+e^{-T_n})
\\ \label{s'P-11}
&<& \alpha_n\eps^3+\beta_n\eps (e^{-T_1}+\cdots+e^{-T_n}).
\end{eqnarray}
Define 
\begin{equation}\label{vP-11}
v_{P^{-1}(1)}=\varphi_{-\tau'_{P^{-1}(1)}}(w_{P^{-1}(1)}).
\end{equation}
Then $v_{P^{-1}(1)}\in\P_{\frac{3\eps}{n}}(x)$ 
and $v_{P^{-1}(1)}=(u'_{P^{-1}(1)},s'_{P^{-1}(1)})_x$ 
satisfy \eqref{u's'L} for $j=P^{-1}(1)$, due to \eqref{u'P-11} and \eqref{s'P-11}.

\noindent
\underline{{Step 4.2:}} {\em Definition of $v_2$.} Denoting $w_2=\varphi_{\hat T_1}(w_{P^{-1}(1)})$
and recalling that $y_2=\Gamma h_2$ is $\hat T_1$-periodic, we apply Lemma \ref{bcbcbc} to write 
\begin{eqnarray*}
w_2&=&\varphi_{\hat T_1}(w_{P^{-1}(1)})
=\Gamma h_2 c_{\bar u_{P^{-1}(1)}+\sigma e^{\hat T_1}} b_{\eta e^{-\hat T_1}}
=\Gamma g c_{u_1}b_{s_1}c_{\sigma_2}b_{\eta_2}c_{\bar u_{P^{-1}(1)}+\sigma e^{\hat T_1}}b_{\eta e^{-\hat T_1}}\\
&=&\Gamma g c_{u'_2}b_{s'_2}a_{\tau'_2}.
\end{eqnarray*}
Analogously, we obtain
\begin{eqnarray} \label{tau2'}
|\tau_2'|&<&\frac{52}{n^2}\,\eps^2,
\\ \notag 
|s_2'-(s_1+\eta_2)|&\leq& \frac{39}{n^3}\,\eps^3+\frac{10}{n}\,\eps e^{-T_1}.
\end{eqnarray}
Therefore it follows from \eqref{eta2-} and \eqref{tildeL-} that
\begin{eqnarray*}
|s_2'-s_2|
\leq |s_2'-(s_1+\eta_2)|+|\eta_2-\tilde s_2|+|\tilde s_2-(s_2-s_1)|
<\frac{77}{n^3}\,\eps^3+\frac{17}{n}\,\eps e^{-T_1},
\end{eqnarray*}
which shows that $s_2'$ satisfies \eqref{sj'L} for $j=2$. 
Furthermore,
$
|u_2'-(u_1+\bar u_{P^{-1}(1)})|
\leq \frac{117}{n^3}\,\eps^3+\frac{8}n \,\eps e^{-T_1}$
yields
\begin{eqnarray*}
|u'_2-u_{P(2)}|
&\leq& 
|u_2'-(u_1+\bar u_{P^{-1}(1)})|+|\bar u_{P^{-1}(1)}-(u_{P(2)}-u_1)|
\\
&\leq &
\Big(\alpha_{n-1}+\frac{217}{n^3}\Big)\eps^3
+\Big(\beta_{n-1}+\frac{17}{n}\Big) (e^{-T_1}+\cdots+e^{-T_n})
\\
&\leq& \alpha_{n}\eps^3+\beta_n\eps (e^{-T_1}+\cdots+e^{-T_n});
\end{eqnarray*}
recall \eqref{baru-}. 
 Hence $u'_2$ satisfies \eqref{uj'L} for $j=2$. 
Defining
\begin{equation*}
\label{v2}
v_2:=\varphi_{-\tau'_2}(w_2)=\Gamma gc_{u'_2}b_{s'_2},
\end{equation*}
we have shown that $v_2\in\P_{\frac{3\eps}{n}}(x)$ and $v_2=(u_2',s_2')$ satisfies \eqref{u's'L}.
\smallskip 
\noindent
\underline{{Step 4.3:}} {\em Definition of $v_j$ with $j\ne P^{-1}(1),j\ne 2$}.
 Since $Q=P_{1,\rm loop}P_1$ is a single-cycle-permutation of the set 
 $\{1,3,\dots,n\}$, we may write
\begin{eqnarray*}
  Q=P_{1,{\rm loop}}P_1  =\big(a,Q(a),Q^2(a),\dots,Q^{n-2}(a)\big),
\end{eqnarray*}
where $a=P^{-1}(1)$. 
For every $j\in\{1,\dots,n\}\setminus \{P^{-1}(1),2\}$, there is $k\in\{1,\dots,n-2\}$
such that $j=Q^k(a)$. Write $z_j=\Gamma \hat h_j$ for $\hat h_j\in\G$ and put
\begin{equation}\label{hatTka}
\hat T_k^a:=\hat T_{P_1Q^0(a)}+\hat T_{P_1Q(a)}+\cdots+\hat T_{P_1Q^{k-1}(a)}.
\end{equation} 
Note that since \eqref{inductive}, we have
$z_j=z_{Q^k(a)}=\varphi_{\hat T^a_k}(z_{P^{-1}(1)})$. 
Set $w_{j}:=\varphi_{\hat T_1+\hat T^a_k}(w_{P^{-1}(1)})$. 
Then using $\Gamma \hat h_j=\Gamma g c_{\hat u_j}b_{\hat s_j}$ and Lemma \ref{bcbcbc}, we write
\begin{eqnarray*}
 w_{j}&=& \Gamma \hat h_{P^{-1}(1)}\,a_{\hat T^a_k}\,
b_{-\bar s_{P^{-1}(1)}e^{-\hat T^a_k}}\,
c_{\sigma e^{\hat T^a_k}}\,b_{\eta e^{-\hat T^a_k}}
=\Gamma\hat h_{j}\,b_{-\bar s_{P^{-1}(1)}e^{-\hat T^a_k}}\,
c_{\sigma e^{\hat T^a_k}}\,b_{\eta e^{-\hat T^a_k}}
\\
&=& \Gamma g c_{\hat u_{j}}b_{\hat s_{j}-\bar s_{P^{-1}(1)}e^{-\hat T^a_k}}\,
c_{\sigma e^{\hat T^a_k}}\,b_{\eta e^{-\hat T^a_k}}\,
= \Gamma g c_{u'_{j}}b_{s'_{j}}a_{\tau'_{j}}
\end{eqnarray*}
and we have
\begin{eqnarray}\label{tau'Qka}
|\tau'_{j}|<\frac{1}{n^2}\,\eps^2,\quad 
  |u'_{j}-\hat u_{j}|<
  \frac{30}{n}\,\eps e^{-T_1},\quad 
  |s'_{j}-\hat s_{j}|
 <\frac{4}{n^3}\,\eps^3. 
\end{eqnarray}
Therefore, using \eqref{hatusj,L}, we obtain
\begin{eqnarray*}
 \notag \label{u'Qka}
  |u'_{j}- u_{P({j})}|
 &\leq& |u'_{j}-\hat u_{j}|
+|\hat u_{j}- u_{P({j})}| 
\\
&\leq & \Big(\alpha_{n-1}+\frac{30}{n^3}\Big)\eps^3+
\frac{30}{n}\,\eps e^{-T_1}+\Big(\beta_{n-1}+\frac{9}{n}\Big)\eps (e^{-T_2}+\cdots e^{-T_n})
\\
&\leq & \alpha_{n}\eps^3+\beta_n \eps(e^{-T_1}+\cdots+e^{-T_n})
\\ \label{s'Qka}
|s'_{j}-s_{j}|
 &\leq&
|s'_{j}-\hat s_{j}|
+|\hat s_{j}-s_{j}|
 \\ 
 &=&\Big(\alpha_{n-1}+\frac{6}{n^3}\Big)\eps^3 +\Big(\beta_{n-1}+\frac{5}{n}\Big)\eps (e^{-T_2}+\cdots e^{-T_n})
 \\
 &\leq & \alpha_n\eps^3+\beta_n\eps (e^{-T_1}+\cdots+e^{-T_n}).
 \end{eqnarray*}
Define
\begin{equation*}\label{vQ^ka}
v_{j}=\varphi_{-\tau'_{j}}(w_{j})=\Gamma g \,c_{u'_{j}}b_{s'_{j}}
\end{equation*}
to obtain $v_j\in\P_{\frac{3\eps}{n}}(x)$; and
\eqref{u'Qka} and \eqref{s'Qka} show that $v_j=(u'_{j},s'_{j})_x$ satisfies \eqref{u's'L}.

\noindent
\underline{{Step 4.4:}} {\em Definition of $T_1',\dots,T_n'$ and proof of (a)}. 
The definition of $w_1,\dots,w_n$ together with \eqref{inductive} lead to
         \begin{equation}\label{wj}
\varphi_{\hat T_j}(w_j)=
\begin{cases} w_{P(j)+1}&\quad\mbox{if}\quad P(j)\ne n\\
w_1&\quad\mbox{if}\quad P(j)=n
\end{cases}
\end{equation}
and $T'=\hat T_1+\cdots+\hat T_n$. Recall 
$v_j=\varphi_{-\tau'_j}(w_j)$, and define  
\begin{equation}\label{TP(j)df}
T_j'=\begin{cases} 
\tau'_{P^{-1}(j)}+\hat T_{j}-\tau'_{j+1}
&\quad \text{if}\quad j\ne n 
\\
\tau'_{P^{-1}(j)}+\hat T_{j}-\tau'_1
&\quad \text{if}\quad j=n.
  \end{cases}
  \end{equation}
Then    $T'=T_1'+\cdots+T_n'$ and 
         \begin{equation}\label{vj}
\varphi_{T'_{P(j)}}(v_j)=
\begin{cases} v_{P(j)+1}&\quad\mbox{if}\quad P(j)\ne n\\
v_1&\quad\mbox{if}\quad P(j)=n.
\end{cases}
\end{equation}

\smallskip 
\underline{{Step 5:}} {\em Proof of (c). }
Recall from
\eqref{tau'p-11}, \eqref{tau2'}, and
\eqref{tau'Qka} that
\quad 
\begin{equation}\label{tau'}
|\tau'_{P^{-1}(1)}|<\frac{1}{n^2}\,\eps^2,\quad 
|\tau_2'|<\frac{52}{n^2}\,\eps^2, \quad
|\tau'_j|< \frac{1}{n^2}\,\eps^2\quad \mbox{for}\quad j\ne 2,\  j\ne P^{-1}(1).
\end{equation}
\underline{Case 1 :} $j=1$. We have
\begin{eqnarray*}
|T_1'-T_1|\leq|T_1'-\hat T_1|+ |\hat T_1- T_1|
\leq |\tau_{P^{-1}(1)}'|+|\tau_2'|+|\hat T_1- T_1|
\leq \frac{82}{n^2}\,\eps^2
< \Delta T_{(n)}
\end{eqnarray*}
due to \eqref{tau'} and \eqref{hatT1-T1}.
\underline{Case 2:} $j\ne 1$.
The observation 
\begin{equation*}
|T_j'-\hat T_j|< \frac{53}{n^2}\,\eps^2\quad \mbox{for all} \quad
j=1,\dots, n
\end{equation*}
yields
\begin{eqnarray*}
|T_j'-T_j|
\leq |T_j'-\hat T_j|+|\hat T_j-\check T_j|+|\check T_j-T_j|
\leq 
\Delta T_{(n-1)}+ \frac{78}{n^2}\,\eps^2 =\Delta T_{(n)}
\end{eqnarray*}
by \eqref{hatTj-checkTj} and \eqref{checkTj-}.

\medskip 
\underline{Step 6:} {\em Proof of (d)}. Note that
$|T'_j-\hat T_j|, |T'_j-T_j|<\Delta T_{(n)}\eps^2<78\eps^2$.
Due to Lemma \ref{lemT-T'}, it remains to show that
\[
d_X(\varphi_t(v_j), \varphi_t(x_{P(j)}))< \frac{40}{n}\, \eps
+\Delta d_{(n-1)}\eps   
\quad \mbox{for all}\quad t\in [0,\min\{T_{P(j)},\hat T_{P(j)}\}].  \] Here the argument is similar to Step 6 in the proof of the case $L=3$. 
We first review some results that will be helpful. Firstly,
recall that
$\tilde x_2=\varphi_{\tilde \tau_2}(x_2)$ and $y_j=\varphi_{-\tilde\tau_2+T_2+\cdots+T_{j-1}}(y_1)$, $j=3,\dots,n$.
Then it follows from \eqref{y1tildex2} and $|\tilde \tau_2|<\frac{8}{n^2}\eps^2$ that
\begin{eqnarray}\label{y1x2,n}
d_X(\varphi_t(y_1),\varphi_t(x_2))< \frac{9}{n}\,\eps+ \frac{8}{ n^2}\,\eps^2
\quad \mbox{for all} \quad t\in [0, T_2]
\end{eqnarray}
and 
\begin{equation}\label{y3x3}
d_X(\varphi_t(y_j),\varphi_t(x_j))<\frac{9}{n}\,\eps\quad
\mbox{for all}\quad t\in [0, T_j+\cdots+T_n], 
\end{equation}  
for $j=3,\dots,n$.  Secondly, due to $w_2=\varphi_{\hat T_1}(w_{P^{-1}(1)})$
and $w_j=\varphi_{\hat T_k^a}(w_2)$ for $j\ne P^{-1}(1), j\ne 2$, 
where $\hat T_k^a$ is defined by \eqref{hatTka}, we have 
\begin{eqnarray}\label{w3chay} \notag 
d_X(\varphi_t(w_j),\varphi_t(z_j))
&=& d_X(\varphi_{t+\hat T_k^a}(w_2), \varphi_{t+\hat T_k^a}(z_{P^{-1}(1)}))
\\ \notag 
&\leq& 
d_X(\varphi_{t+\hat T_k^a}(w_2), \varphi_{t+\hat T_k^a}(\bar z_{P^{-1}(1)})) +d_X(\varphi_{t+\hat T_k^a}(\bar z_{P^{-1}(1)} ),
\varphi_{t+\hat T_k^a}(z_{P^{-1}(1)} ))
\\ \notag 
&\leq& \frac{30}{n}\,\eps +|\bar\tau_{P^{-1}(1)}|
<\frac{31}{n}\,\eps\quad \mbox{for all}\quad t\in [0,\hat T_{P(j)}],
\end{eqnarray}
using \eqref{inductive} and \eqref{phitwp-112}; recall $\bar z_{P^{-1}(1)}=\varphi_{-\bar \tau_{P^{-1}(1)}}(z_{P^{-1}(1)})$. 

\noindent
\underline{Step 6.1:} $j=P^{-1}(1)$. 
For $t\in [0,\min\{\hat T_1,\tilde T_1\}]$, 
\begin{eqnarray*}
d_X(\varphi_t(v_{P^{-1}(1)}),\varphi_t(x_1))
&\leq & d_X(\varphi_t(v_{P^{-1}(1)}),\varphi_t(w_{P^{-1}(1)}))
+d_X(\varphi_t(w_{P^{-1}(1)}),\varphi_t(y_2))
+d_X(\varphi_t(y_2),\varphi_t(x_1))
\\
&<&|\tau_{P^{-1}(1)}'|
+\frac{30}n\,\eps+\frac{9}{n}\,\eps
<\frac{39}{n}\,\eps+\frac{1}{n^2}\,\eps^2,
\end{eqnarray*}
due to \eqref{tau'p-11}, \eqref{phitwp-11}, and \eqref{x1y2}.
Since $|\hat T_1-\tilde T_1|<\frac{21}{n^2}\eps^2$, we obtain
\begin{eqnarray*}
d_X(\varphi_t(v_{P^{-1}(1)}),\varphi_t(x_1))<
\frac{39}{n}\,\eps +\frac{1}{n^2}\,\eps^2+\sqrt 2\cdot\frac{21}{n^2}\,\eps^2<
\frac{40}{n}\,\eps
\quad\mbox{for all}\quad t\in [0, \hat T_1].
\end{eqnarray*}
\noindent
\underline{Step 6.2:} $j=2$. Recall that $|\tau_2'|<\frac{52}{n^2}\eps^2,\,
|\bar\tau_{P^{-1}(1)}|<\frac{44}{n^2}\eps^2,
|\check\tau_{P(2)}|<\frac{1}{n^2}\eps^2$. \underline{Case 1:} $P(2)\ne 2$. For $t\in[0,\min\{T_{P(2)},\hat T_{P(2)}\}]$, we have
\begin{eqnarray*}
&&\hskip-1cm{d_X(\varphi_t(v_2),\varphi_t(x_{P(2)}))}
\\
&\leq& \,\,
d_X(\varphi_t(v_2),\varphi_t(w_2))
+d_X(\varphi_t(w_2),\varphi_t(\bar z_{P^{-1}(1)}))
+d_X(\varphi_t(\bar z_{P^{-1}(1)}),\varphi_t(z_{P^{-1}(1)}))
\\
&&+\,\,d_X(\varphi_t( z_{P^{-1}(1)}),\varphi_t(\check y_{P(2)}))+d_X(\varphi_t(\check y_{P(2)}), \varphi_t(y_{P(2)}))
+d_X(\varphi_t(y_{P(2)}),\varphi_t(x_{P(2)}))
\\
 &\leq&|\tau_2'|+|\bar\tau_{P^{-1}(1)}|
 +|\check\tau_{P(2)}|
 +d_X(\varphi_{t+\hat T_1}(w_{P^{-1}(1)}), \varphi_t(\bar z_{P^{-1}(1)}))
 \\
 && +\ d_X(\varphi_t(z_{P^{-1}(1)}),\varphi_t(\check y_{P(2)}))
+d_X(\varphi_t(y_{P(2)}),\varphi_t(x_{P(2)}))
\\
&<& \frac{97}{n^2}\,\eps^2+\frac{30}n\,\eps+\Delta d_{(n-1)}\eps +\frac{9}{n}\,\eps 
<\frac{40}{n}\,\eps+\Delta d_{(n-1)}\eps,
\end{eqnarray*}
using \eqref{phitwp-112}, \eqref{d(n-1)}, and \eqref{y3x3}.
\underline{Case 2:} $P(2)=2$. Analogously to Case 1, by replacing $y_{P(2)}, \check y_{P(2)}$, and
$\check\tau_{P(2)}$ by $y_1,\check y_1$, and $\check \tau_1$, respectively, and using \eqref{y1x2,n}, we obtain 
\begin{eqnarray}
d_X(\varphi_t(v_2),\varphi(x_{P(2)}))
<\frac{113}{ n^2}\,\eps^2 + \frac{39}{n}\, \eps  +\Delta d_{(n-1)}\eps<\frac{40}{n}\,\eps +\Delta d_{(n-1)}\eps, 
\end{eqnarray}
recalling $|\check \tau_1|<\frac{16}{n^2}\eps^2$ from \eqref{checktau1}.
\underline{Step 6.3:} $j\ne P^{-1}(1), j\ne 2$.  \underline{Case 1:} $P(j)\ne 2$. For $t\in [0, \min\{T_{P(j)},\hat T_{P(j)}\}]$, using \eqref{w3chay}, 
\eqref{d(n-1)}, and \eqref{y3x3}, we obtain  
\begin{eqnarray*}
d_X(\varphi_t(v_j),\varphi_t(x_{P(j)}))
&\leq& d_X(\varphi_t(v_j),\varphi_t(w_j))+
d_X(\varphi_t(w_j),\varphi_t( z_j))
+d_X(\varphi_t(z_j),\varphi_t(\check y_{P_1(j)}))\\
&&+\,\,d_X(\varphi_t(\check y_{P_1(j)}),\varphi_t(y_{P_1(j)}))+
d_X(\varphi_t(y_{P_1(j)}),\varphi_t(x_{P_1(j)}))
\\ 
&<&|\tau_j'|+|\check\tau_{P(j)}|
+\frac{31}n\,\eps +\Delta d_{(n-1)}\eps+\frac{9}{n}\,\eps
\\&<&\frac{40}{n}\,\eps+\Delta d_{(n-1)}\eps;
\end{eqnarray*}
note that $|\tau_j'|<\frac{52}{n^2}\eps^2$ by \eqref{tau'}
and $|\check\tau_j|<\frac{16}{n^2}\eps^2$ by \eqref{checktau1} and \eqref{checktauj}.   \underline{Case 2:} $P(j)=2$. Here the argument is similar to Case 2 in Step 6.2.

\medskip 

\underline{Step 7:} {\em We prove that $|u_j'|<\frac{3}{n}\eps, |s_j'|<\frac{3}{n}\eps$.} By Step 5, it has been shown that 
\begin{equation}\label{vj-xpj}
d_X(\varphi_t(v_j),\varphi_t(x_{P(j)}))<\Delta d_{(n)}\eps<\eps_* \quad\mbox{for all}\quad t\in [0,\max\{T_{P(j)},T'_{P(j)}\}]
\end{equation}
for $j=1,\dots,n$; where $\eps_0=\eps(\rho)$ is from Theorem \ref{eps0} with respect to $\rho=1$
and $\eps_*$ is from Lemma \ref{per-coinc}. Then   $\eps_*<\eps_0$ implies that
  $|u_j'-u_{P(j)}|<e^{-T_{{P(j)}}}<\frac{2}{n}\,\eps$
as well as 
$|u'_{j}|<\frac{3}{n}\eps$.
Furthermore,
recall \eqref{vj} and  $\varphi_{T_1}(x_1)=x_2$, $\varphi_{T_2}(x_2)=x_3,\dots$, $\varphi_{T_n}(x_n)=x_1$. It follows from \eqref{vj-xpj} that 
\begin{eqnarray*}
d_X( \varphi_t(v_{P(j)+1}),\varphi_t(x_{P(j)+1}))<\eps_0\quad \mbox {for all}\quad t\in\big[-\max\{T_{P(j)},T'_{P(j)}\},0\big]
\end{eqnarray*}
for $j=1,\dots,n$ with the convention $P(j)+1\equiv 1$ if $P(j)=n$.
 This means that 
\begin{equation}\notag
d_X(\varphi_t(v_j),\varphi_t(x_j))<\eps_0\quad\mbox{for all}\quad 
t\in \big[-\max\{T_{j-1},T'_{j-1}\},0\big].
\end{equation}
Then by Theorem \ref{eps0},
$
|s_j'-s_j|<|s_j's_j||u_j'-u_j|+e^{-T_{j-1}}$
leads to
\[|s_j'|<\frac{|s_j|+e^{-T_{j-1}}}{1-|s_j||u_j'-u_j|}<\frac{\frac{\eps}{n}
+\frac{5\eps}{3n}}{1-\frac{\eps}{n}\cdot\frac{4\eps}{n}}<\frac{9}{8}\cdot \frac{8}{3n}\,\eps=\frac{3}n\,\eps.\]

\smallskip 
\underline{Step 8:} {\em The distinction between $c$ and $c'$}.  Recall from Step 7 that
\begin{equation}
\label{uj'-upj}
|u_j'-u_{P(j)}|< e^{-T_{P(j)}}
\end{equation}
and 
\begin{equation}\label{sj'-sj}
|s_j'-s_j|<|s_j's_j||u_j'-u_j|+e^{-T_{j-1}}<\frac{3}{L^2}\,\eps^2\cdot\frac{4}{L}\,\eps+e^{-T_{j-1}}=\frac{12}{L^3}\,\eps^3+e^{-T_{j-1}}
\end{equation}
for $j=1,\dots,L$. By assumption and what we have shown, 
$c$ and $c'$ cross the Poincar\'e section of $x$ at $x_j=(u_j,s_j)_{x}$ and
$v_j=(u_j',s_j')_x,\, j=1,\dots,L$, respectively. If $c$ and $c'$ do coincide, then
 they will have the same intersections with the Poincar\'e section at $x$. 
 For any $i,j\in\N_L:=\{1,\dots,L\}$ given,  we prove that $v_j\ne x_i$.
\underline{Case 1:} $P(j)\ne i$. Then by condition (ii) and \eqref{uj'-upj} it follows that
\begin{eqnarray*}
|u_j'-u_i|&\geq& |u_i-u_{P(j)}|-|u_{P(j)}-u_j'|
>\frac65\,(e^{-T_{P(j)}}+e^{-T_i})-e^{-T_{P(j)}}
\\
&=&\frac{1}{5}\,e^{-T_{P(j)}}+\frac{6 }{5}\,e^{-T_i}>0
\end{eqnarray*} 
and hence $u_j'\ne u_i$ as well as $v_j\ne x_i$. \underline{Case 2:} $P(j)=i\ne j$.
 We prove that $s_j'\ne s_i$. Indeed, due to condition (ii) and \eqref{sj'-sj}, we have
\begin{eqnarray*}
|s_j'-s_i|&\geq& |s_j-s_{P(j)}|-|s_j-s_j'|
>\frac{24}{L^3}\,\eps^3+e^{-T_{j-1}}+e^{-T_{P(j)-1}} -\Big(\frac{12}{L^3}\,\eps^3+ e^{-T_{j-1}}\Big)
\\
&=&\frac{12}{L^3}\,\eps^3+e^{-T_{{P(j)}-1}}>0
\end{eqnarray*}
so that $s_j'\ne s_i$ as well as $v_j\ne x_i$. \underline{Case 3:} $P(j)=i=j$.
 For a contradiction, we suppose that $v_j=x_j$. Then the orbits $c$ and $c'$
  do agree and they have the same intersections with $\P_\eps(x)$. 
  Denote by $F_P$ the set of fixed points of $P$. 
For any $k\in F_P$, we show that $v_k=x_k$. Indeed, suppose that $v_{k_0}\ne x_{k_0}$ for some $k_0\in F_P$. 
This implies that $v_{k_0}=x_{i_0}$ for some $i_0\ne k_0$. Then $P(k_0)=k_0\ne i_0$ which contradicts Case 1.
 Therefore $v_k=x_k$ for all $k\in F_P$. Since $P$ is not the identity permutation, there is $k_0\in \N_L$
 so that $P(k_0)\ne k_0$.
  By above, $v_{k_0}=x_{l_0}$ for some $l_0 \in \N_L, P(l_0)\ne l_0$.  
If $P(k_0)\ne l_0$ we have a contradiction to Case 1. 
If $P(k_0)=l_0$ then we have a contradiction to Case 2. Therefore $v_j\ne x_j$ as was to be shown. 
{\hfill$\Box$}\bigskip

\begin{remark}\rm
In order to show that $c$ has $(L-1)!-1$ partners, we have to make sure the partner orbits are pairwise distinct. 
Let $Q$ be a permutation in $S_L$ such that $P_{\rm loop}Q$ is a single cycle and $Q\ne P$. 
By Theorem \ref{thmL}, the orbit $c$ has a partner orbit $c''\ne c$ which is described by the permutation $Q$.
 Furthermore, there are $L$ points $w_1,\dots,w_L$ in $c''$ such that $w_j\in\P_\eps(x)$ with 
$w_j=(u_j'',s_j'')_x$ satisfying 
\begin{eqnarray}\label{uj'-uQj}
|u_j''-u_{Q(j)}|<e^{-T_{Q(j)}},\quad
|s_j''-s_j|< \frac{12}{L^3}\,\eps^3+e^{-T_{j-1}}
\end{eqnarray}
for all $j=1,\dots,L$ with the convention $T_{j-1}=T_L$ for $j=1$. 
Then $c''$ is different from $c$. Now we prove that $c'$ and $c''$ do not coincide. 
Fixing $i,j\in\N_L$, we show that $w_j\ne v_i$. \underline{Case 1:} $Q(j)\ne P(i)$.
 Then due to condition (ii), \eqref{uj'-upj}, and \eqref{uj'-uQj} we have
\begin{eqnarray}\notag
|u_j''-u_i'|
&\geq&  |u_{Q(j)}-u_{P(i)}|-|u_j''-u_{Q(j)}|-|u_i'-u_{P(i)}|
\\ \notag
&>&\frac{6}5\,(e^{-T_{Q(j)}}+e^{-T_{P(i)}})- e^{-T_{Q(j)}}-e^{-T_{P(i)}}>0;
\end{eqnarray}
hence $u_j''\ne u_i'$ and as a consequence $w_j\ne v_i$. 
\underline{Case 2:} $Q(j)=P(i)$ and  $j\ne i$.  Then
\begin{eqnarray*}
|s_j''-s_i'|
&\geq& |s_j-s_i|
-|s''_i-s_i|-
|s'_i-s_i|
\\
&>& \frac{24}{L^3}\,\eps^3+ e^{-T_{j-1}} +e^{-T_{i-1}}-\Big(\frac{12}{L^3}\,\eps^3 +e^{-T_{j-1}}\Big)
-\Big(\frac{12}{L^3}\,\eps^3 +e^{-T_{i-1}}\Big)
=0
\end{eqnarray*}
implies that $s_j''\ne s_i$ and thus $w_j\ne v_i$. 
\underline{Case 3:} $Q(j)=P(i)$ and $j=i$. For a contradiction, suppose that $w_j=v_i$. 
Then the orbits $c'$ and $c''$ do agree, so that they have the same intersections with $\P_\eps(x)$. 
Denote by $I$ the subset of $\N_L$ consisting of $k$ so that $Q(k)=P(k)$.
Note that $I$ may be empty but $I\ne \N_L$ since $ Q\ne P$. 
If $I\ne \emptyset$ then $w_k=v_k$ for all $k\in I$ since otherwise 
 $w_{k_0}=v_{l_0}$  for some $k_0\in I$ and $l_0\ne I$ implies $Q(l_0)\ne P(k_0)$ which contradicts Case 1.
 Fix $k,k'\in \N_L\setminus I$ such that $w_k=v_{k'}$. 
 According to Case 1, we have $Q(k)=P(k')$. This implies that $k\ne k'$ since $k,k'\ne I$. Then 
 a contradiction to Case 2 is obtained, and therefore $w_j\ne v_i$. 
{\hfill$\diamondsuit$}
\end{remark}

\begin{remark} \rm
(a)
If $P(j)=j$ for some $j\in \{1,\dots,L\}$ then we count the $L$-encounter as an $(L-1)$-encounter,
 and the error in the estimate of action difference would be better. 
 
 \smallskip
 \noindent (b) For the case $L=3$, since the given orbit has a unique partner orbit, condition (ii) in Theorem \ref{3-encounter}   can be reduced  to 
$|u_1-u_2|>e^{-T_2}, |u_1-u_3|>e^{-T_3}$, and $|s_1-s_3|>\frac{4}{9}\eps^3 +e^{-T_2}$. 
Due to Theorem \ref{eps0}, Condition (ii) in Theorem \ref{thmL}  makes sure that any two encounter stretches  are not too close
for the whole time before intersecting the Poincar\'e section again, i.e., 
they do not simultaneously approach a shorter  periodic orbit. In other words, the stretches are separated by non-vanishing loops (also called intervening loops or the stretches do not overlap), also needed in physics literatures, see \cite{muellerthesis,mueller2005}.

\smallskip
\noindent (c) If a periodic orbit has several encounters, for example one $L$-encounter and one $N$-encounter, $L,N\geq 3$, the partner orbits can be constructed as the following. We apply Theorem \ref{thmL} for the $L$-encounter to have a new periodic orbit. This orbit has a $N$-encounter whose entrance ports and exit ports are like the ones of the original orbit. Then we can apply Theorem \ref{thmL} for this encounter and we will obtain the corresponding partner for the original one.
{\hfill$\diamondsuit$}
\end{remark}

\end{document}